\documentclass[twocolumn]{aastex62}

\turnoffedit

\usepackage{amsmath}
\usepackage[flushleft]{threeparttable} 
\usepackage{array}
\newcommand{\kms}{km~s$^{-1}$}

\newcommand{\co}[1][]{\ensuremath{^{#1}}CO}

\let\vec\mathbf

\defcitealias{Boyden16}{B16}

\shorttitle{Feedback in Orion A}
\shortauthors{Feddersen {\em et al.}}

\begin{document}

\title{The CARMA-NRO Orion Survey: Statistical Signatures of Feedback in the Orion A Molecular Cloud}

\author{Jesse R. Feddersen}
\affiliation{Department of Astronomy, Yale University, P.O. Box 208101, New Haven, CT 06520-8101, USA}
\author{H\'ector G. Arce}
\affiliation{Department of Astronomy, Yale University, P.O. Box 208101, New Haven, CT 06520-8101, USA}
\author{Shuo Kong}
\affiliation{Department of Astronomy, Yale University, P.O. Box 208101, New Haven, CT 06520-8101, USA}
\author{Volker Ossenkopf-Okada}
\affiliation{  I.\,Physikalisches Institut, Universit\"at zu K\"oln,  
  Z\"ulpicher Stra{\ss}e 77,  50937 K\"oln, Germany}
\author{John M. Carpenter}
\affiliation{Joint ALMA Observatory, Alonso de Cordova 3107 Vitacura, Santiago de Chile, Chile}

\email{jesse.feddersen@yale.edu}

\begin{abstract}
We investigate the relationship between turbulence and feedback in the Orion A molecular cloud using maps of \co[12](1-0), \co[13](1-0) and C$^{18}$O(1-0) from the CARMA-NRO Orion survey. We compare gas statistics with the impact of feedback in different parts of the cloud to test whether feedback changes the structure and kinematics of molecular gas. We use principal component analysis, the spectral correlation function, and the spatial power spectrum to characterize the cloud. We quantify the impact of feedback with momentum injection rates of protostellar outflows and wind-blown shells as well as the surface density of young stars. We find no correlation between shells or outflows and any of the gas statistics. However, we find a significant anti-correlation between young star surface density and the slope of the \co[12] spectral correlation function, suggesting that feedback may influence this statistic. While calculating the principal components, we find peaks in the covariance matrix of our molecular line maps offset by 1-3 \kms~toward several regions of the cloud which may be produced by feedback. We compare these results to predictions from molecular cloud simulations. 
\end{abstract}


\section{Introduction}
Stars form deep inside giant molecular clouds (GMCs) \citep{McKee07,Heyer15}. Young stars output mechanical and thermal energy, or feedback, into their birth clouds \citep{Krumholz14}. This feedback may decrease the efficiency of star formation by counteracting gravitational collapse \citep{Federrath15} and help drive and maintain turbulence \citep{Nakamura07,Offner18}.

GMCs are turbulent, with supersonic linewidths that increase with physical size \citep{Zuckerman74,Larson81,Mac-Low04}. However, turbulence decays rapidly \citep{Mac-Low98, Stone98,Padoan99}, and must be maintained by some mechanism. Feedback from young stars may help maintain the turbulence of molecular clouds \citep{Li06,Matzner07,Carroll09}.

\citet{Offner18} proposed a mechanism for translating local feedback into large scale turbulent driving. They used magnetohydrodynamic simulations to show that feedback effects may be propagated through a cloud by magnetic fields. Upon injecting winds into the simulation, they showed that the velocity dispersion outside of the wind-blown shells was increased and the velocity power spectrum flattened. These effects were caused by magnetosonic waves coming from the compressed wind-blown shells. These magnetosonic waves could explain how feedback drives turbulence at larger scales.

Feedback in molecular clouds has mostly been studied by cataloging individual features such as protostellar outflows \citep[e.g.,][]{Arce10, Plunkett15}, photon-dominated regions (PDRs) with far-UV heating and photoablation \citep[e.g.,][]{Bally18}, and stellar wind-blown shells \citep[e.g.,][]{Arce11, Nakamura12, Li15, Feddersen18}. The physical characteristics of these features can then be measured to estimate the impact of feedback on a molecular cloud. However, visually cataloging feedback features is time-consuming and prone to significant bias and difficulty of separating features from the rest of the cloud.

Recently, several studies have considered the impact of feedback by measuring statistics of molecular gas structure and motions. \citet{Nakamura07} and \citet{Carroll09} showed that the presence of outflows modifies the velocity power spectrum of simulated molecular clouds, producing peaks at the scale where outflows inject energy into the cloud. \citet{Padoan09} investigated the power spectrum of the molecular cloud NGC 1333, finding no evidence for a departure from a power-law near the outflow energy injection scales predicted by the above-mentioned simulations.

\citet{Swift08} computed the power spectrum of the red and blue line wings of \co[13] in the L1551 molecular cloud, which hosts several outflows. They found a feature in these power spectra at a scale of about 0.05~pc, indicating a preferential scale of energy injection into the cloud. \citet{Sun06} compared the CO power spectrum in different regions of the Perseus molecular cloud. They found the most actively star-forming region NGC 1333 had a steeper slope than the quiescent dark cloud L1455. \citet{Swift08} also measured a flat linewidth-size relationship in the low-mass molecular cloud L1551, suggesting that turbulent motions originate to a large degree at small-spatial scales. This is contrary to the turbulent cascade that is usually assumed for
molecular clouds, where driving happens at large scales and dissipation occurs at the smallest scales via gas viscosity \citep{Mac-Low04}.

Principal component analysis \citep{Heyer97} has also been used to investigate the effect of feedback on molecular clouds. \citet{Brunt09} showed that the ratio between different-order principal components of a simulated cloud is sensitive to the driving scale of turbulence. Adding outflows to their simulations did not change this ratio, implying that feedback in the form of outflows was not driving turbulence in their simulated cloud. However, \citet{Carroll10} showed that this analysis is biased towards the largest scales. The presence of significant smaller-scale turbulent driving from feedback may then be hidden in the principal component analysis. 

The spectral correlation function quantifies the similarity of pairs of spectra as a function of their separation. \citet{Ballesteros-Paredes02} applied the spectral correlation function to 21 cm HI spectra of the North Celestial Pole Loop region. Instead of averaging the spectral correlation function (see Equation~\ref{eq:scf} below) over all pixels, they constructed a map of the local SCF of each pixel. Their map of the SCF highlighted the edge of the expanding supernova remnant HI shell. They suggested using the SCF as a tool for finding shells.

\citet{Boyden16} (hereafter \citetalias{Boyden16}) applied the TurbuStat\footnote{http://turbustat.readthedocs.io/en/latest/} \citep{Koch17} suite of statistical measures to molecular cloud simulations from~\citet{Offner15}. These simulations tested the effect of stellar winds on the structure of a simulated molecular cloud and successfully reproduced the expanding shells found in Perseus by \citet{Arce11}. \citetalias{Boyden16} found statistical measures sensitive to the mass-loss rate of the injected stellar winds. In their simulation with winds, the covariance matrix (see Section~\ref{sec:methods_pca}) showed peaks of spatially correlated emission separated by 1-3~\kms~and the spectral correlation function (see Section~\ref{sec:methods_scf}) had a steeper slope compared to simulations without winds. They found the power spectrum was insensitive to stellar winds.

In this paper, we test whether statistical measures of CO in the Orion A GMC trace feedback in the cloud, as predicted by \citetalias{Boyden16}. In Section~\ref{sec:data}, we describe the CARMA-NRO Orion observations and split the cloud into subregions. In Section~\ref{sec:quantifying_feedback}, we define three ways of quantifying feedback in the cloud. In Section~\ref{sec:methods}, we introduce the statistics we use to summarize the CO data. In Section~\ref{sec:results}, we present the results of these statistics and compare with previous studies. In Section~\ref{sec:discussion}, we discuss the relationship between the statistics and feedback in the cloud. In Section~\ref{sec:conclusions}, we summarize the conclusions of the paper and suggest future directions for the statistical study of feedback.

\begin{figure*}
\plotone{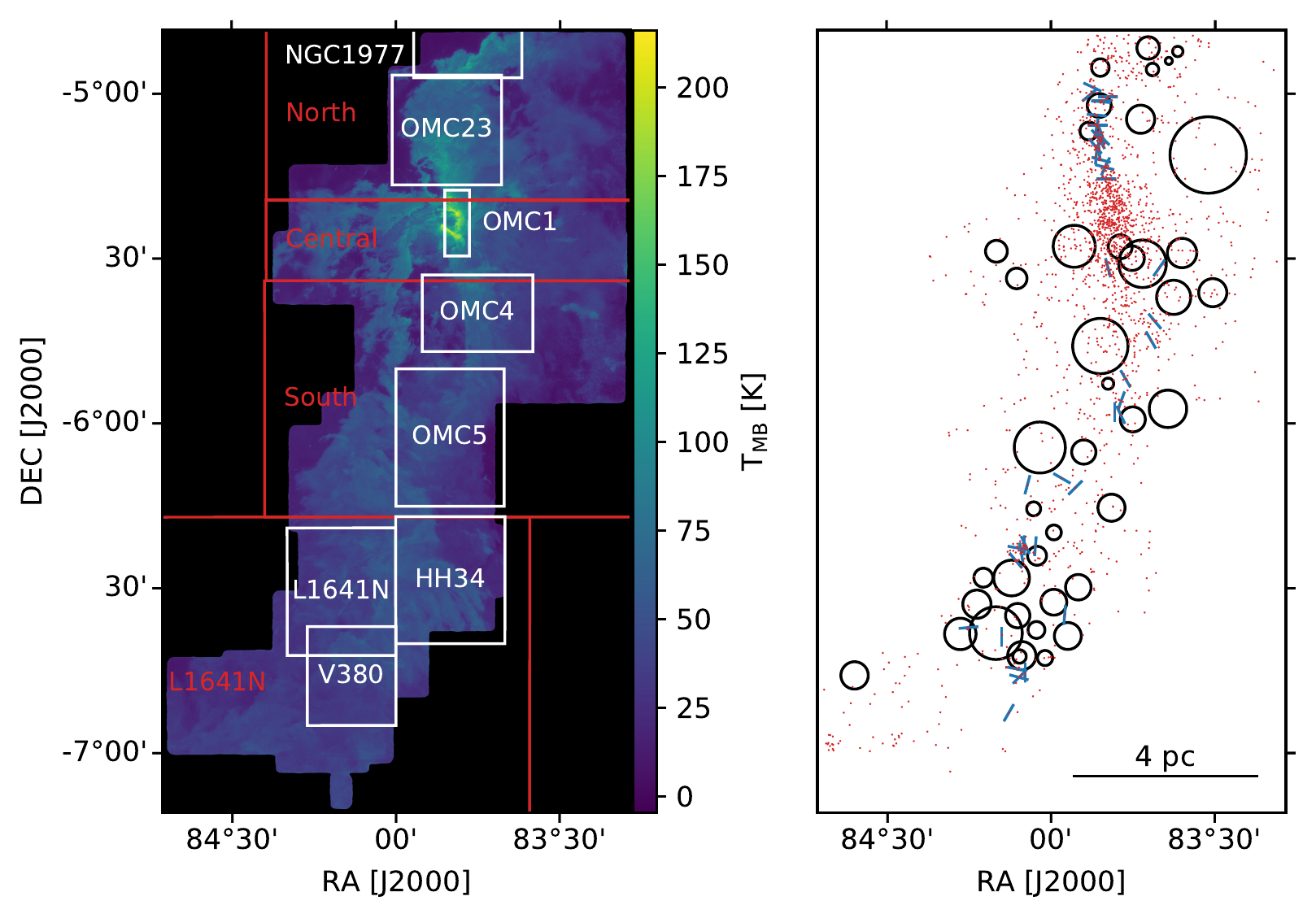}
\caption{
Finding charts of subregions and sources of feedback used in this paper.
\emph{Left}: Subregions overlaid on the CARMA-NRO Orion \co[12] peak temperature map. The small subregions outlined in white come from \citet{Davis09} except for OMC1, which comes from \citet{Ungerechts97}. The large subregions outlined in red come from \citet{Feddersen18}. \emph{Right}: The three measures of feedback impact are shown. The \emph{black circles} show the location and radius of the candidate expanding shells from \citet{Feddersen18}. The \emph{blue lines} show the location and orientation of outflows cataloged by \citet{Tanabe:submitted}. The length of the lines is arbitrary and does not indicate outflow length. The \emph{red points} mark all pre-main sequence stars and protostars from the Spitzer Orion catalog of \citet{Megeath12}. \label{fig:overview}}
\end{figure*}

\section{Data}\label{sec:data}
\subsection{CARMA-NRO Orion CO Maps}
We use the \co[12](1-0), \co[13](1-0), and C$^{18}$O(1-0) spectral-line maps of Orion A from the CARMA-NRO Orion Survey \citep{Kong18}. These maps were obtained by combining interferometric images from the Combined Array for Millimeter/Submillimeter Astronomy (CARMA) with single-dish maps from the Nobeyama Radio Observatory (NRO) 45m radio telescope. This method preserves the angular resolution of CARMA while also recovering large-scale structure. The combined maps probe physical scales of 0.01 - 10 pc at a distance of 414~pc \citep{Menten07}.\footnote{The part of Orion A covered by our survey has been located about 5\% closer (at about 380-400~pc) with GAIA Data Release 2 parallaxes \citep{Kounkel18,Grossschedl18,Kuhn19}. Adopting a distance of 390~pc would decrease the projected distances and the derived physical quantities (momentum injection rates) in this paper by a factor of 0.94. Our conclusions are independent of distance.} The \co[12] and C$^{18}$O maps have a beam full-width at half-maximum (FWHM) of $10\arcsec\times8\arcsec$ and the \co[13] map has a FWHM of $8\arcsec\times6\arcsec$. The \co[12] velocity resolution is 0.25~\kms~while \co[13] and C$^{18}$O have a velocity resolution of 0.22~\kms. For more details on the CARMA-NRO Orion data, see \citet{Kong18}.

\subsection{Subregions}
To compare the statistics of turbulence with feedback impact, we would ideally like to have a control - a cloud which is identical in every way to Orion A except with no stellar feedback. In this ideal case, any statistical differences between the clouds could be attributed solely to the action of feedback. This ideal scenario was simulated by \citetalias{Boyden16} but no such true control cloud exists for Orion A, though \citet{Lada09} proposed that the California Molecular Cloud is an Orion A analogue with an order of magnitude lower star formation rate. In this study, we compare different regions of Orion A with different amounts of feedback, assume that this is the only relevant difference, and look for trends in the statistical methods that resemble those found in the simulated clouds of \citetalias{Boyden16}. A similar approach is used by \citet{Sun06} in their comparison of power spectra in different regions of the Perseus molecular cloud. 

We divide the Orion maps into several subregions, guided by previous studies. \citet{Feddersen18} split Orion A into four subregions - North, Central, South, and L1641N - to compare the impact of expanding shells with protostellar outflows and cloud turbulence. The North subregion includes the NGC 1977 PDR \citep{Peterson08}, OMC-2/3, and the M43 HII region. The Central subregion covers a wide variety of environments, including the Orion Bar PDR \citep{Goicoechea16}, OMC-1 (the densest part of the cloud), the explosive Orion BN/KL outflow \citep{Bally17}, and more diffuse gas to the east and west. The South subregion covers OMC 4/5 and the pillar-shaped PDRs to the east dubbed the dark lane south filament (DLSF) by \citet{Shimajiri11}. The L1641N subregion covers the low-mass cluster L1641-N \citep{Nakamura12} and the reflection nebula NGC 1999 \citep{Stanke10}.

In addition to these regions, we consider several smaller subregions focused on individual parts of the cloud
based on the molecular hydrogen outflow survey of \citet{Davis09}. These smaller subregions focus on specific clusters or groups of young stars in the Orion A cloud and all have similar projected areas on the sky of about 0.1~deg$^2$ (5~pc$^2$). We also add a subregion defined by \citet{Ungerechts97} which is restricted to the densest core of Orion A - OMC 1, BN-KL, and the Orion Bar. We define these subregions following conventional definitions in Orion A to avoid cherry-picking regions that happen to show the correlations between statistics and feedback that we are looking for. Figure~\ref{fig:overview} shows the subregions on a map of \co[12] peak temperature and Table~\ref{tab:regions} defines the extent of each subregion.

\begin{deluxetable*}{cccccccc}[htb!]
\tablecaption{\label{tab:regions}Subregion Results}
\tablehead{\colhead{Name}&\colhead{Right Ascension (J2000)}&\colhead{Declination (J2000)}&\colhead{$\dot P_{\rm shell}$\tablenotemark{d}}&\colhead{$\dot P_{\rm out}$\tablenotemark{e}}&\colhead{$\log_{10} n_{\rm YSO}$\tablenotemark{f}}&\colhead{$^{12}$CO $\alpha_{\rm SCF}$}&\colhead{$^{12}$CO $\alpha_{\rm SPS}$}
}
\startdata
North\tablenotemark{a} & $5^{\rm h}37^{\rm m}35^{\rm s}$ to $5^{\rm h}33^{\rm m}06^{\rm s}$ & $-5\arcdeg19\arcmin26\arcsec$ to $-4\arcdeg48\arcmin19\arcsec$ & 17 [6, 41] & $6.2 \pm 0.5$ & 3.2 & $-0.096 \pm 0.004$ & $-3.07 \pm 0.04$ \\ 
Central\tablenotemark{a} & $5^{\rm h}37^{\rm m}35^{\rm s}$ to $5^{\rm h}33^{\rm m}06^{\rm s}$ & $-5\arcdeg34\arcmin03\arcsec$ to $-5\arcdeg19\arcmin21\arcsec$ & 33 [14, 64] & $0.13 \pm 0.04$ & 3.4 & $-0.139 \pm 0.002$ & $-3.1\pm 0.06$ \\ 
South\tablenotemark{a} & $5^{\rm h}37^{\rm m}36^{\rm s}$ to $5^{\rm h}33^{\rm m}07^{\rm s}$ & $-6\arcdeg17\arcmin03\arcsec$ to $-5\arcdeg34\arcmin03\arcsec$ & 41 [19, 77] & $0.7 \pm 0.2$ & 2.9 & $-0.087 \pm 0.006$ & $-3.43 \pm 0.03$ \\ 
L1641N\tablenotemark{a} & $5^{\rm h}39^{\rm m}01^{\rm s}$ to $5^{\rm h}34^{\rm m}22^{\rm s}$ & $-7\arcdeg11\arcmin44\arcsec$ to $-6\arcdeg17\arcmin03\arcsec$ & 24 [10, 52] & $2.2 \pm 0.3$ & 2.7 & $-0.059 \pm 0.002$ & $-3.0 \pm 0.03$ \\ 
NGC1977\tablenotemark{b} & $5^{\rm h}35^{\rm m}47^{\rm s}$ to $5^{\rm h}34^{\rm m}28^{\rm s}$ & $-4\arcdeg57\arcmin08\arcsec$ to $-4\arcdeg37\arcmin23\arcsec$ & 33 [14, 64] & $0$ & 3.3 & $-0.121 \pm 0.006$ & $-3.6 \pm 0.14$ \\ 
OMC23\tablenotemark{b} & $5^{\rm h}36^{\rm m}03^{\rm s}$ to $5^{\rm h}34^{\rm m}43^{\rm s}$ & $-5\arcdeg16\arcmin37\arcsec$ to $-4\arcdeg56\arcmin37\arcsec$ & 28 [4, 85] & $19 \pm 1$ & 3.4 & $-0.138 \pm 0.003$ & $-3.42 \pm 0.07$ \\ 
OMC1\tablenotemark{c} & $5^{\rm h}35^{\rm m}24^{\rm s}$ to $5^{\rm h}35^{\rm m}06^{\rm s}$ & $-5\arcdeg29\arcmin35\arcsec$ to $-5\arcdeg17\arcmin35\arcsec$ & 93 [33, 186] & $0$ & 4.2 & $-0.158 \pm 0.001$ & $-3.83 \pm 0.29$ \\ 
OMC4\tablenotemark{b} & $5^{\rm h}35^{\rm m}41^{\rm s}$ to $5^{\rm h}34^{\rm m}20^{\rm s}$ & $-5\arcdeg46\arcmin56\arcsec$ to $-5\arcdeg33\arcmin00\arcsec$ & 162 [78, 279] & $0.19 \pm 0.05$ & 3.4 & $-0.111 \pm 0.002$ & $-4.46 \pm 0.19$ \\ 
OMC5\tablenotemark{b} & $5^{\rm h}36^{\rm m}00^{\rm s}$ to $5^{\rm h}34^{\rm m}41^{\rm s}$ & $-6\arcdeg15\arcmin07\arcsec$ to $-5\arcdeg50\arcmin06\arcsec$ & 18 [6, 41] & $2.9 \pm 0.6$ & 2.9 & $-0.079 \pm 0.002$ & $-4.04 \pm 0.1$ \\ 
HH34\tablenotemark{b} & $5^{\rm h}36^{\rm m}00^{\rm s}$ to $5^{\rm h}34^{\rm m}40^{\rm s}$ & $-6\arcdeg40\arcmin07\arcsec$ to $-6\arcdeg17\arcmin00\arcsec$ & 23 [13, 40] & $0.18 \pm 0.04$ & 2.7 & $-0.082 \pm 0.002$ & $-4.18 \pm 0.11$ \\ 
L1641N\tablenotemark{b} & $5^{\rm h}37^{\rm m}20^{\rm s}$ to $5^{\rm h}36^{\rm m}00^{\rm s}$ & $-6\arcdeg42\arcmin15\arcsec$ to $-6\arcdeg19\arcmin02\arcsec$ & 36 [15, 71] & $5.0 \pm 0.9$ & 2.8 & $-0.077 \pm 0.002$ & $-4.04 \pm 0.1$ \\ 
V380\tablenotemark{b} & $5^{\rm h}37^{\rm m}05^{\rm s}$ to $5^{\rm h}36^{\rm m}00^{\rm s}$ & $-6\arcdeg55\arcmin00\arcsec$ to $-6\arcdeg37\arcmin00\arcsec$ & 22 [4, 75] & $7 \pm 1$ & 2.7 & $-0.070 \pm 0.002$ & $-4.22 \pm 0.12$ \\ 
\enddata
\tablenotetext{a}{Subregions from \citet{Feddersen18}}
\tablenotetext{b}{Subregions from \citet{Davis09}}
\tablenotetext{c}{Subregions from \citet{Ungerechts97}}
\tablenotetext{d}{Momentum injection rate surface density of shells. Units are $10^{-3}$ M$_{\odot}$ km s$^{-1}$ yr$^{-1}$ deg$^{-2}$. Values in brackets are summed lower and upper limits from Table 3 of \citet{Feddersen18}.}
\tablenotetext{e}{Momentum injection rate surface density of outflows. Units are $10^{-3}$ M$_{\odot}$ km s$^{-1}$ yr$^{-1}$ deg$^{-2}$. Uncertainties are calculated by adding in quadrature the individual outflow uncertainties in Table 7 of \citet{Tanabe:submitted}.}
\tablenotetext{f}{Surface density of YSOs. Units are deg$^{-2}$. Because of catalog incompleteness, this is likely an underestimate in the Central and OMC1 regions.}
\end{deluxetable*}

\section{Quantifying Feedback}\label{sec:quantifying_feedback}
In order to relate gas statistics to the impact of feedback in the cloud, we attempt to measure this impact. 
\citetalias{Boyden16} quantified feedback using the mass-loss rate of stellar winds injected into their simulations. \citet{Offner15} designed these simulations to reproduce the expanding shells observed by \citet{Arce11} in the Perseus molecular cloud. Thus, to make the closest comparison to \citetalias{Boyden16}, we first consider the expanding shells in Orion A identified by \citet{Feddersen18}.

\subsection{Expanding Shells}\label{sec:shells}
\citet{Feddersen18} identified 42 expanding shells in Orion A using the CO maps from the NRO 45m telescope. The authors visually identified expanding structures in the CO channel maps and matched many of these structures with low- and intermediate-mass young stars. Similar shells have been found in the Perseus \citep{Arce11} and Taurus \citep{Li15} molecular clouds. While the origin of such shells is unclear, one explanation is spherical stellar winds from young stars which entrain the cloud material into expanding shells. We show the shells of \citet{Feddersen18} in Figure~\ref{fig:overview}.

The impact of shells on the cloud is summarized by Table 3 in \citet{Feddersen18}. In this study, we quantify a shell's impact on the cloud by its momentum injection rate ($\dot P_{\rm shell}$). The total shell momentum injection rate in each subregion is the sum of $\dot P_{\rm shell}$ for each shell centered inside that subregion. The true shell impact is uncertain, as it is difficult to extract shell emission cleanly from the rest of the cloud. Furthermore, most shells are incomplete. They are only detected over a fraction of the expected volume in the spectral cube. Lower and upper limits on $\dot P_{\rm shell}$ are estimated in Table 3 of \citet{Feddersen18} and can span a factor of several above or below the median value. We ignore these lower and upper limits and instead focus on the relative shell impacts between subregions, but the uncertainties mean any results based on the shell momenta are inconclusive. Because of these large uncertainties and the potential for false-positive bias in identifying expanding shells, we use independent methods to quantify feedback.

\subsection{Protostellar Outflows}\label{sec:outflows}
Young stars that are actively accreting material launch collimated bipolar outflows which impact the surrounding cloud material and may help drive or maintain turbulence \citep{Arce07,Frank14}. In Orion A, outflows have been identified by many authors \citep[e.g.][]{Morgan91,Williams03,Stanke07,Takahashi08}. A comprehensive census of outflows found in our NRO 45m data has been carried out by \citet{Tanabe:submitted}. They have detected about 50\% more outflows than previously known. Notably, they identified 11 outflows in the poorly studied OMC 4/5 region where none were previously known. \citet{Tanabe:submitted} excluded the area around OMC-1 from their search because the YSOs in this region are crowded and the cloud velocity width is too broad to disentangle outflow emission. Therefore, the outflow measurements for our regions around OMC1 are incomplete. 

To quantify the impact of outflows on the cloud, we use the momentum injection rates ($\dot P_{\rm out}$) tabulated in Table 6 of \citet{Tanabe:submitted}. Many of the outflows in this table have multiple lobes listed separately. In these cases, we sum the individual lobes of each outflow. We assign outflows to the subregion containing the driving source in Table 3 of \citet{Tanabe:submitted}. The only case where the driving source of an outflow lies in a different subregion from part of the outflow emission is Outflow 19 in \citet{Tanabe:submitted}. A small portion of the emission from this outflow falls south of the OMC23 subregion, where its driving source is located. In every other case, the outflow emission and its driving source are in the same subregion.

To measure the outflow mass, \citet{Tanabe:submitted} first integrate the \co[12] emission. \co[12] is often optically thick, which means line intensity is no longer directly proportional to column density and mass. To account for the optical depth of \co[12], other tracers must be observed. For example, \citet{ZhangY16} use \co[13] and C$^{18}$O combined with \co[12] to measure outflow masses more accurately. \citet{Tanabe:submitted} only detect a few outflows in both \co[12] and \co[13]. They calculate the average optical depth of these few outflows and apply this correction factor to every outflow in their catalog, ignoring any variation in optical depth.

After they find the mass of an outflow, \citet{Tanabe:submitted} they calculate  momentum by multiplying this mass by the line-of-sight velocity of the outflow. But if the outflow axis is inclined relative to the line-of-sight, this will underestimate the true momentum. 

Variable optical depth and inclination angle both introduce uncertainty in the outflow momentum injection rates. The optical depth of outflows in Orion A is likely not uniform. This fact is evidenced by the different outflow optical depths measured by \citet{Tanabe:submitted} as well as the variations in the ratio between the integrated intensities of \co[12] and \co[13] throughout the cloud \citep[see Figure 25 in][]{Kong18}. A variable outflow optical depth will affect our comparisons of outflow impact between subregions.

\subsection{Young Stars}\label{sec:ysos}
As an independent estimate of feedback we consider the young stars (YSOs) in each subregion. YSOs are more directly a measure of star formation rate than feedback impact. However, YSOs are ultimately responsible for both outflows and shells. Therefore we consider the surface density of YSOs $n_{\rm YSO}$ to be a proxy for the relative strength of feedback in different regions.

To measure $n_{\rm YSO}$, we use the Spitzer Orion catalog of YSOs \citep{Megeath12}. They classified stars as protostars or pre-main sequence stars with disks on the basis of their mid-IR colors. In this classification, 86\% of the Spitzer Orion YSOs are pre-main sequence stars and the remaining 14\% are protostars. Shells are likely driven by pre-main sequence stars \citep{Arce11,Feddersen18} while outflows are more likely driven by protostars. Therefore, we calculate $n_{\rm YSO}$ using all YSOs in the catalog.

Because the YSO catalog is a mixture of more evolved pre-main sequence stars and younger protostars, this measure traces a wider range of timescales than either shells or outflows alone. As noted in Section~5.2.3 of \citet{Arce10}, the cumulative impact of feedback in a cloud over the course of star formation may be greater than what is traced by the currently active outflows and shells. Thus, if there are YSOs which have ejected outflows or powered shells no longer detectable as coherent structures in the cloud, then $n_{\rm YSO}$ may be a better tracer of the potential link between feedback and turbulence than shells or outflows alone.

To compare subregions of different sizes, we calculate the surface density of each feedback measure, dividing by the projected area of each subregion. The Spitzer Orion catalog suffers from incompleteness toward regions with bright IR nebulosity \citep{Megeath16}. Thus the YSO surface density in the Central and OMC1 subregions is likely higher relative to the surface density in other subregions. From Figure 2 and 3 in \citet{Megeath16}, the typical completeness fraction in the ONC is approximately 0.5. Therefore, the true central YSO surface density may be up to about twice the value reported here. Our three feedback measures are the momentum injection rate surface density of shells/outflows (M$_\odot$~km~s$^{-1}$~yr$^{-1}$~deg$^{-2}$) and the YSO surface density (deg$^{-2}$).

\section{Statistical Methods}\label{sec:methods}
We use the TurbuStat\footnote{https://turbustat.readthedocs.io/en/latest} package, described in detail by \citet{Koch17}, to compute the principal component analysis, spectral correlation function, and spatial power spectrum of the \co[12], \co[13], and C$^{18}$O data in Orion A. \citetalias{Boyden16} also found several other statistics that were sensitive to wind mass loss rate (see their Table 3). However, \citet{Boyden18} incorporated astrochemical models into these simulations and found that many of the statistics were also sensitive to the chemical complexity, radiation field, and molecular tracer used. The statistics which are least sensitive to these effects while remaining sensitive to feedback strength were principal component analysis and the spectral correlation function (see Section~5.3 in \citealt{Boyden18}). While the spatial power spectrum was not sensitive to feedback in \citetalias{Boyden16}, its form has been studied for possible signatures of feedback \citep[e.g.][]{Swift08} so we include it here. TurbuStat also provides a distance metric for each statistic, which allows comparison of different cubes. We briefly describe these statistical methods below.

\subsection{Principal Component Analysis}\label{sec:methods_pca}
Principal component analysis (PCA) is a statistical technique used to reduce the dimensionality of a dataset. \citet{Ungerechts97} and \citet{Heyer97} first applied PCA to molecular line maps to study the chemistry and turbulence in molecular clouds.

The method implemented by TurbuStat to compute the PCA of a spectral cube comes from \citet{Heyer97}. A spectral cube with $n$ pixels can be expressed as a set of $n$ spectra with a number $p$ of velocity channels. This can be represented as a matrix $T(r_i, v_j) \equiv T_{ij}$ with $n$ rows and $p$ columns, where $r_i$ is the position of pixel $i$ and $v_j$ is the velocity of channel $j$. 

First, each element of the covariance matrix $S$ is calculated to be

\begin{equation}\label{eq:cov}
S_{jk} = S(v_j, v_k) = \frac{1}{n} \sum^n_{i=1}T_{ij}T_{ik}.
\end{equation}

This covariance matrix is then diagonalized to find its eigenvalues and eigenvectors. Projecting the spectral cube onto each eigenvector gives a set of eigenimages called principal components. Each eigenvalue correspond to the amount of variance recovered by each principal component.

\citetalias{Boyden16} found the covariance matrix to be sensitive to the injected stellar winds. Their Figure 3 shows peaks at 1-3~\kms~in the covariance matrix of the simulation with winds which are absent in the simulation without winds. We test whether such features appear in the Orion covariance matrices.

To compare their simulated spectral cubes, \citetalias{Boyden16} used the Turbustat PCA distance metric. Each set of eigenvalues for a particular cube is sorted in descending order; then the eigenvalues are normalized by dividing them by their sum. The PCA distance metric between two cubes is then the Euclidean distance between the two normalized sets of eigenvalues, or the square root of the sum of the square differences. \citetalias{Boyden16} found a strong correlation between the PCA distance metric and winds. \citet{Boyden18} incorporated gas chemistry into these simulations and found that PCA also varied between models with and without chemistry. However, the covariance peaks remained a unique signature of winds. For more details on PCA, see \citet{Brunt02,Brunt02a,Brunt13}. 

\subsection{Spectral Correlation Function}\label{sec:methods_scf}
The spectral correlation function (SCF) was first introduced by \citet{Rosolowsky99} and refined by \citet{Padoan01}. The SCF measures the similarity of spectra as a function of their spatial separation, or lag. The SCF at a specific lag vector $\Delta \vec{r}$ (between two pixels) is defined to be

\begin{equation}\label{eq:scf}
\mathrm{SCF}(\vec{r}, \Delta \vec{r}) = 1 - \sqrt{\frac{\sum_v [T(\vec{r}, v) - T(\vec{r} + \Delta \vec{r}, v)]^2}{\sum_v T(\vec{r}, v)^2 + \sum_v T(\vec{r} + \Delta \vec{r}, v)^2}},
\end{equation}
where $\vec{r}$ is the position of a pixel, $\Delta \vec{r}$ is the lag vector, $v$ is velocity, and $T$ is the temperature (or intensity). The SCF is then averaged over all pixels $\vec{r}$. Repeating this for various lag vectors, a 2D spectral correlation surface can be constructed. An azimuthal average of this surface, or equivalently an average over all rotated lag vectors of the same length, is a 1D spectrum of the SCF as a function of spatial separation. We refer to this 1D spectrum as the SCF in this paper. \citet{Padoan01} showed that the SCF is well characterized by a power-law in both simulated and observed molecular clouds, over a wide range of physical scales. They also showed that the slope of the SCF is independent of velocity resolution and signal-to-noise.

The SCF distance metric is defined in TurbuStat as the sum of the square differences between the SCF surfaces, weighted by the inverse square distance from the center of the surface. \citetalias{Boyden16} found SCF to be sensitive to the strength of feedback. In their simulations with winds injected, the SCF has a significantly steeper slope than in the wind-free simulations. More recently, \citet{Boyden18} found that including chemistry in these simulations flattened the SCF slope as opposed to the steepening seen with winds. For more details on SCF, see \citet{Rosolowsky99} and \citet{Padoan01}.

\subsection{Spatial Power Spectrum}\label{sec:methods_sps}
The power spectrum is the Fourier transform of the two-point autocorrelation (or square) of the integrated intensity. We azimuthally average this two-dimensional power spectrum to arrive at a one-dimensional power spectrum which we hereafter refer to as the SPS. See \citet{Stutzki98} for the general $n$-dimensional derivation and \citet{Pingel18} for a detailed description of the SPS implementation used here.

Because our maps have emission at the edges, the Fourier transform is affected by strong ringing \citep[e.g.][]{Brault71,Muller04} which can be seen at small spatial scales in the power spectrum. To correct for this ringing, we taper the integrated intensity with a Tukey window where the outer 20\% of the map is gradually reduced to zero. This tapering also reduces the noise at the edges of the observed maps. Additionally, the observed beam introduces artificial correlation into the map \citep[e.g.][]{Dickey01}. To correct for this, we divide the power spectrum of the integrated intensity by the power spectrum of the ellipsoidal Gaussian beam. This introduces a divergence at very small scales but allows us to extend the dynamic range over which the underlying power spectrum is recovered. These corrections are implemented in TurbuStat and described in tutorials included in the package documentation.

\citetalias{Boyden16} found that the SPS was not sensitive to feedback in their simulations of winds, while \citet{Boyden18} found that temperature variation flattened the slope. However, some studies have suggested that feedback induces a break in the power spectrum. \citet{Swift08} reported a ``bump'' in the \co[13] SPS of L1551 at a scale of about 0.05~pc. They attributed this peak to the energy injection scale of the outflows in the cloud. However, they used the power spectrum of the line wings which may be more sensitive to outflow emission than the full integrated intensity of the cloud. We discuss the line wing power spectra in Section~\ref{sec:discussion_linewing}.

\section{Results}\label{sec:results}

\subsection{Principal Component Analysis}
\begin{figure*}
\plotone{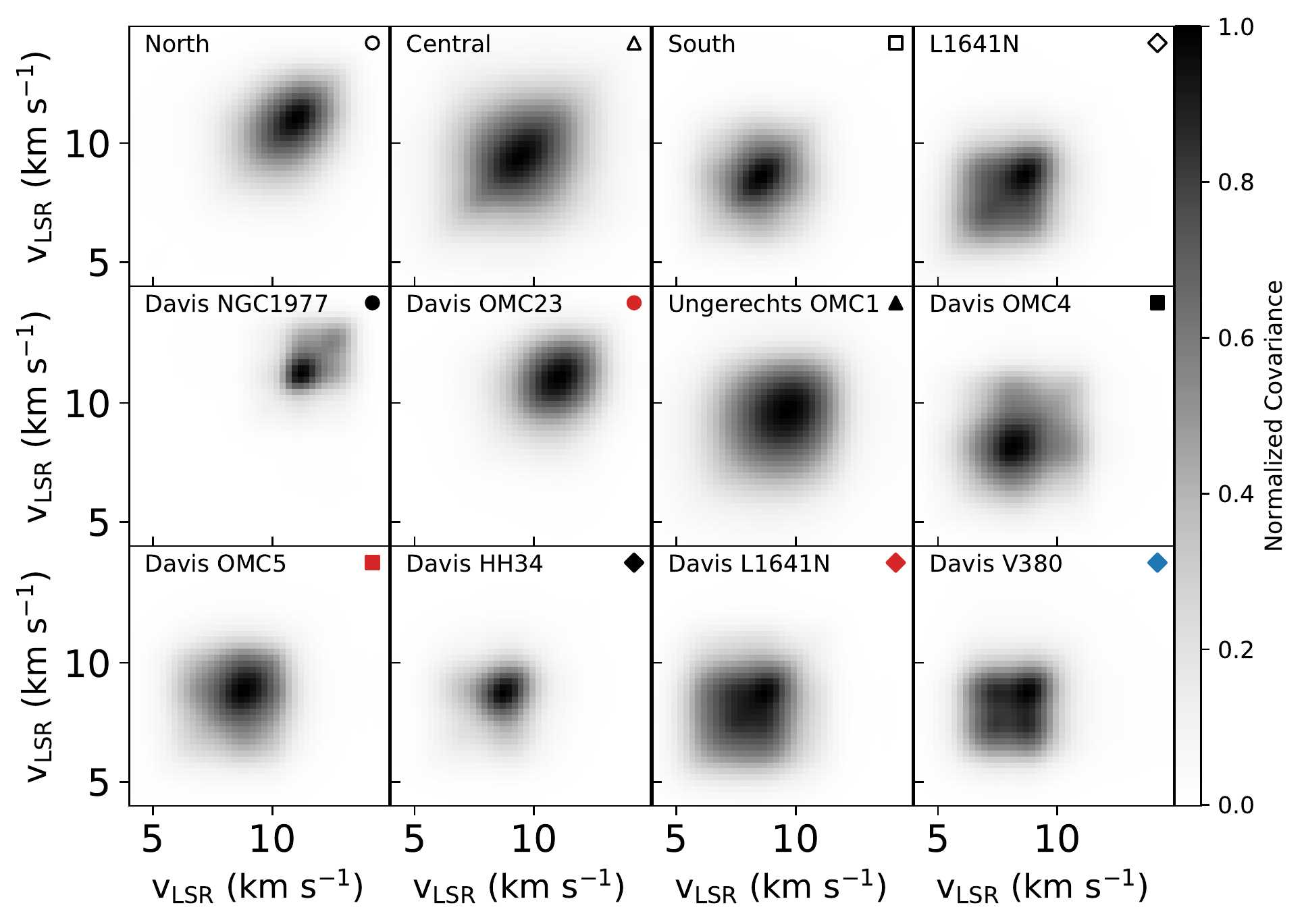}
\caption{\co[12] covariance matrices in Orion A subregions. The top row shows the large subregions used by \citet{Feddersen18}, ordered from north to south. The bottom two rows show the smaller subregions defined by \citet{Ungerechts97} and \citet{Davis09}, ordered from north to south. The symbol in each panel is used in subsequent figures to denote each subregion. The covariance matrix is used for the eigenvalue decomposition that forms the basis of principal component analysis, and is symmetric about the one-to-one axis by construction. Each matrix is normalized by its maximum covariance. Features offset from the one-to-one axis by 1 - 3 \kms~are seen most prominently in the V380, NGC1977, OMC4, and L1641N subregions and resemble the features \citetalias{Boyden16} noticed in their simulations of winds.\label{fig:cov_12co}}
\end{figure*}

\citetalias{Boyden16} showed that the covariance matrix (Equation~\ref{eq:cov}), an intermediate product of PCA on spectral cubes, was sensitive to the presence of feedback. In their simulations with winds included, the covariance matrix shows several peaks at $1-3$~\kms~which are not present in the simulations without winds. In Figure~\ref{fig:cov_12co}, we show the covariance matrices of the Orion A \co[12] subregions (the \co[13] and C$^{18}$O covariance matrices are described in Appendix~\ref{sec:appendix_13co_c18o}). In several regions, most prominently NGC 1977, L1641N, OMC 4, and V380, we find covariance peaks offset from the diagonal axis by $1-3$~\kms~as in the \citetalias{Boyden16} wind simulations. Essentially, these covariance peaks mean there is emission in these subregions which is spatially correlated but separated by a few \kms~in velocity. This spatially correlated (but velocity-offset) emission can clearly be seen in the channel maps of these regions \citep{Kong18}. The source of these features could be feedback, as it is in the \citetalias{Boyden16} simulations, or some other effect. 

\citet{Nakamura12} proposed that the L1641N cluster is located at the intersection of two colliding clouds. This idea is based on the overlapping velocity components in this region. The blue velocity component (4-6~\kms) dominates emission to the southeast of L1641N while the red component (7-12~\kms) extends north of the cluster. \citet{Nakamura12} noticed these velocity components in channel maps, but these are the same velocity components that appear as peaks in our covariance matrices of the L1641N and V380 regions. It is unclear whether the covariance peaks are the result of this cloud collision scenario or the expansion of wind-blown shells like those simulated by \citetalias{Boyden16} or some combination of the two effects. 

In the NGC 1977 region, the molecular cloud is excited by FUV radiation from the H II region to the north \citep{Kutner85,Makinen85}. This FUV may be responsible for photoablation~\citep{Ryutov03} of the northern edge of the molecular cloud, accelerating it to a few~\kms \/ and generating the covariance peaks seen in this region. Well known examples of photoablative flows can be found in the Orion Bar~\citep{Goicoechea16} and the Horsehead Nebula~\citep{Bally18}. Detailed comparison of the covariance matrix in both simulated and observed molecular clouds is needed to fully understand the mechanisms behind these peaks.

\citetalias{Boyden16} used the PCA distance metric to further show that PCA was sensitive to the winds in their simulations (see their Figure 16). We find no correlation between any of our feedback measures and the PCA distance metric between subregions. This could be because our feedback measures are not good proxies for the winds simulated by \citetalias{Boyden16}, or because another mechanism unrelated to feedback is driving the velocity structure traced by the covariance peaks. 

\subsection{Spectral Correlation Function}

\begin{figure*}
\plotone{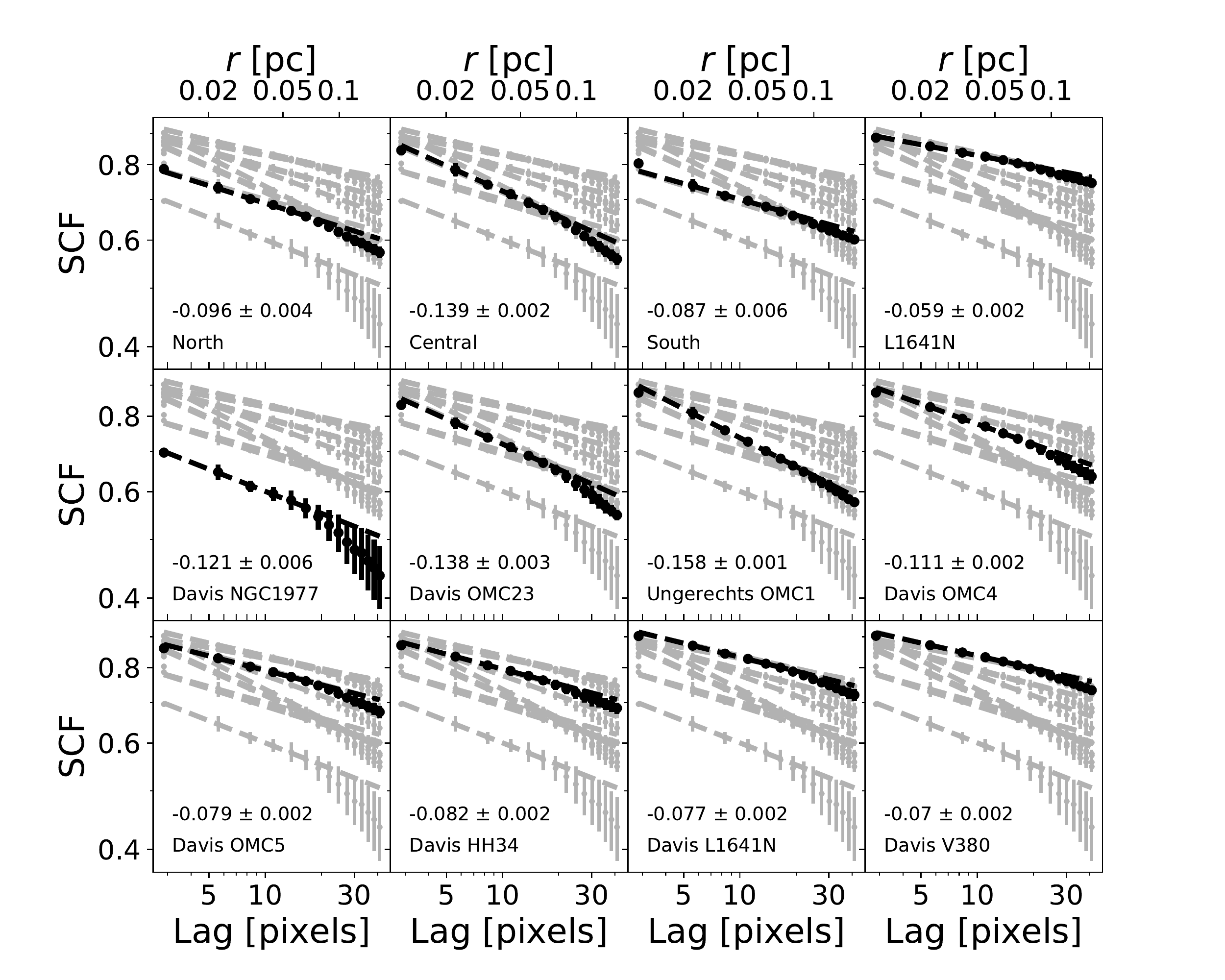}
\caption{\co[12] spectral correlation function (SCF) in Orion A subregions. The subregions are ordered in the same way as Figure~\ref{fig:cov_12co}. In each panel, the \emph{black points} are the azimuthally averaged values of the SCF surface, with errors given by the standard deviation of the SCF surface in each bin. The \emph{black line} is a weighted least-squares power law fit to these points. The \emph{gray lines and points} show the SCF and fits for all subregions. Each SCF is fit between lags of 5 to 17 pixels. The power law slope and uncertainty are shown in each panel. \label{fig:scf_12co}}
\end{figure*}

Figure~\ref{fig:scf_12co} shows the \co[12] SCF and power-law fits of each subregion. The fit slopes are tabulated in Table~\ref{tab:regions}. We compute the two-dimensional SCF surface at lags between 0 and 30 pixels (0 to 0.12 pc) in intervals of 3 pixels (0.01 pc). Using a lag interval of 1 pixel does not significantly change the resulting SCF, so we save computational time by only calculating the SCF surface every 3 pixels. We then average the SCF surface in equally-spaced annuli to arrive at the one-dimensional SCF spectrum shown in~Figure~\ref{fig:scf_12co}. We calculate the SCF over the same range of spatial scales as shown in Figure 6 of \citetalias{Boyden16} for the most direct comparison. Each subregion's SCF follows a power-law closely up to a lag of approximately 20 pixels. We compute a weighted least-squares fit to each SCF between lags of 5 to 17 pixels (approximately 0.02 to 0.07~pc). Unlike in the \citetalias{Boyden16} simulations, the Orion SCF steepens at larger lags. \citet{Gaches15} also found a steepening SCF in \co[13] maps of Ophiuchus and Perseus. We describe the SCF at scales larger than 0.12~pc in Appendix~\ref{sec:appendix_scf_largescale}.

The SCF slopes in the Orion A subregions range from -0.15 to -0.06. \citet{Gaches15} calculated the \co[12] SCF of the Perseus and Ophiuchus molecular clouds and found slopes around -2, steeper than in any of our Orion A subregions. However, limited by the angular resolution of their data they fit the SCF at larger scales: 0.1 - 1 pc. \citet{Padoan03} found \co[13] SCF slopes between -0.1 and -0.5 in various molecular clouds. Most of their SCF spectra are fit at larger scales ($> 0.1$ pc) than those presented here. But the smallest maps in their dataset, L1512 and L134a, have similar spatial resolution to our Orion A maps and also have the shallowest SCF slopes at -0.18 and -0.13, respectively. The simulations in \citetalias{Boyden16} also have steeper SCF slopes compared to our data. The SCF is well known to vary with spatial resolution \citep{Gaches15} making comparison between different datasets difficult. \citetalias{Boyden16} found the SCF slope was sensitive to their simulated wind mass loss rate. In Section~\ref{sec:discussion_regions} we compare the SCF slope between subregions and discuss the relationship between SCF slope and feedback impact.

\subsection{Spatial Power Spectrum}
\begin{figure*}
\plotone{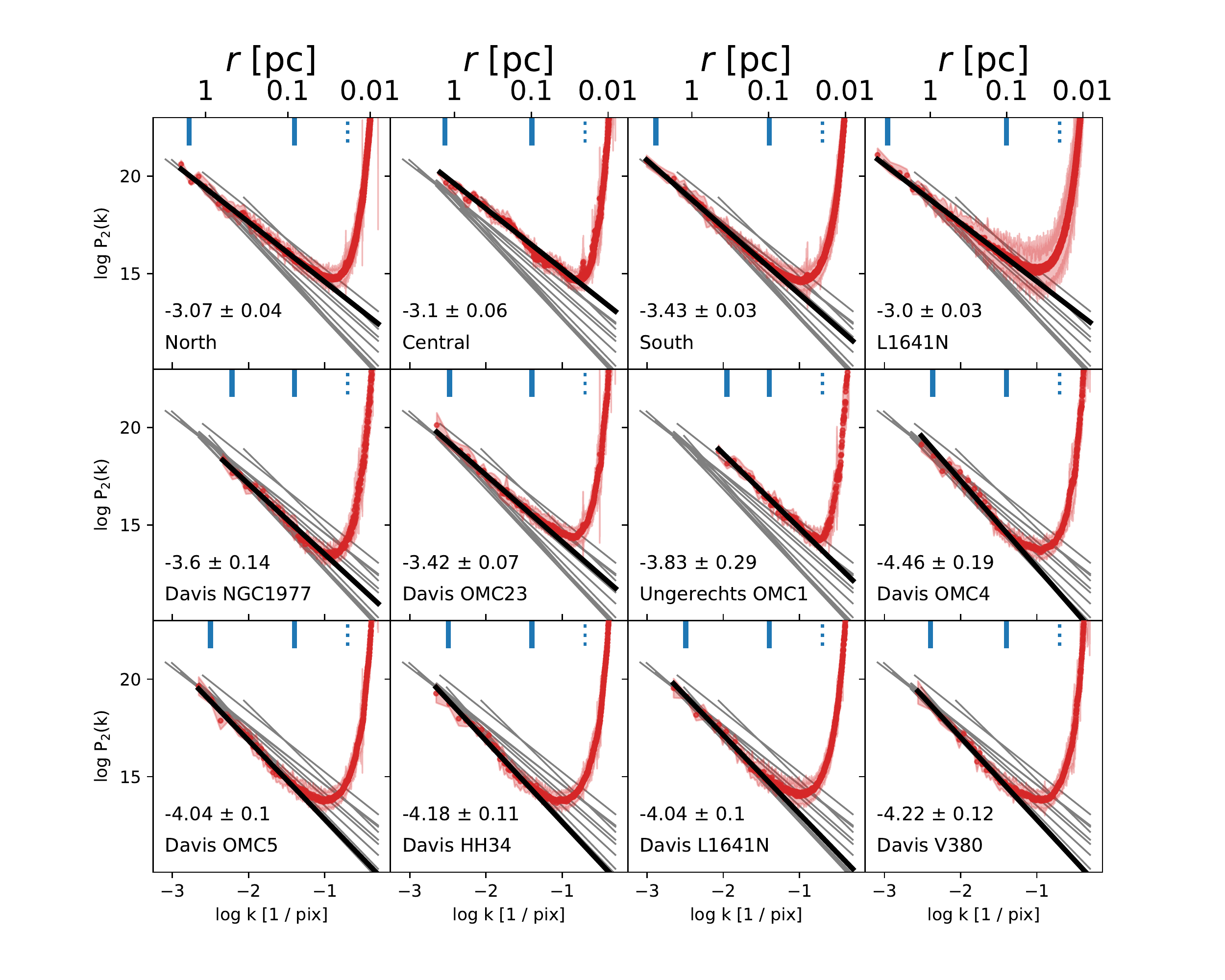}
\caption{\co[12] spatial power spectrum (SPS) in Orion A subregions. The subregions are ordered in the same way as Figure~\ref{fig:cov_12co}. In each panel, the \emph{red points and red shaded region} are the values and uncertainties of the SPS. The \emph{black line} is a weighted least-squares power law fit to these points. The \emph{solid blue vertical lines} at the top of each panel show the range over which the power law is fit. The \emph{dotted blue vertical line} at the top of each panel shows the FWHM of the beam's major axis. The \emph{gray lines} show the SPS fits for all subregions. Each power spectrum is fit down to a scale of -1.4 (25 pixels), or about five times the beam FWHM. Each fit also excludes the largest scale point in the power spectrum. The power-law slope and its uncertainty is shown in each panel.\label{fig:sps_12co}}
\end{figure*}

Figure~\ref{fig:sps_12co} shows the \co[12] power spectra and power-law fits for each subregion. The fit slopes are tabulated in Table~\ref{tab:regions}. The beam correction described in Section~\ref{sec:methods_sps} causes the sharp upturn at scales smaller than about twice the beamwidth. We restrict the power-law fits to spatial scales greater than about five times the beamwidth. We also exclude the largest scale point in each power spectrum from the fits as some of the power spectra show slight deviations at this largest scale. In the fitted regime, the power spectra closely follow a power-law with no evidence of the peaks seen in \citet{Swift08}. 

The power-law fits to the SPS in the Orion A subregions have slopes ranging between about -3 and -4 in \co[12] and \co[13] with somewhat shallower slopes between about -2.2 and -3.4 in C$^{18}$O (see Appendix~\ref{sec:appendix_13co_c18o} for \co[13] and C$^{18}$O SPS). The SPS slope of an optically thick medium is predicted to saturate to -3 for a wide range of physical conditions (such as sound speed and magnetic field strength) \citep{Lazarian04,Burkhart13}. Previous studies of the \co[12] and \co[13] SPS in molecular clouds have shown slopes close to this optically thick limit of -3 \citep[][e.g.]{Stutzki98,Padoan06,Sun06,Pingel18}, shallower than our SPS slopes. However, these studies measure the SPS of entire clouds instead of smaller regions within clouds, averaging over larger areas than our subregions. \citet{Sun06} reported the power spectrum slope in several regions within the Perseus Molecular Cloud spanning $50\arcmin \times 50\arcmin$, or $4.4~\mathrm{pc} \times 4.4~\mathrm{pc}$ at a distance of 300 pc \citep{Zucker18,Ortiz-Leon18}. Most of these regions have SPS slopes steeper than -3, more similar to our subregions than to cloud-wide power spectra. Also, our \emph{largest} subregions have slopes closest to the theoretically predicted value of -3. In Section~\ref{sec:discussion_regions}, we compare the SPS slope to feedback impact in each subregion.

\section{Discussion}\label{sec:discussion}
\subsection{Spectral Slopes and Feedback Impact}\label{sec:discussion_regions}

In the simulations of \citetalias{Boyden16}, SCF responds strongly to the strength of feedback but is independent of evolutionary time and magnetic field strength. They found that SCF slope steepened with increased wind mass-loss rate. \citet{Boyden18} showed that SCF is sensitive to gas chemistry, but including chemistry flattens the SCF slope - the opposite impact of feedback. Thus, SCF may still probe the relative strength of feedback in different regions, especially if chemistry is similar between regions.

SPS, on the other hand, has only a weak dependence on feedback in \citetalias{Boyden16} but varies strongly with evolutionary time and magnetic field strength, making SPS a poor diagnostic for feedback. Further, \citet{Boyden18} showed that SPS is also sensitive to temperature variations in their simulations which include chemistry.

\begin{figure*}
\plotone{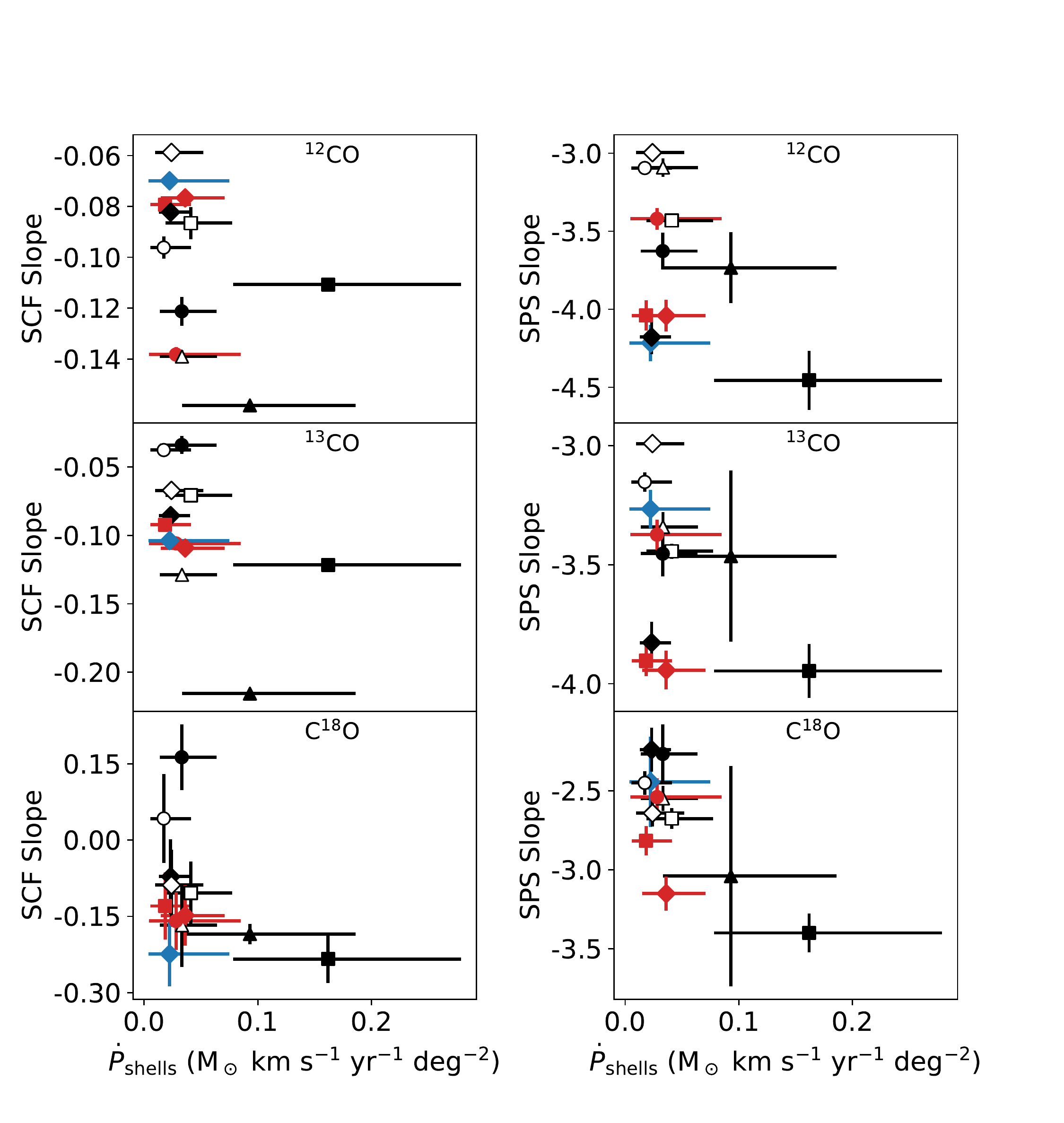}
\caption{The SCF and SPS slopes versus the  momentum injection rate surface density of shells in Orion A subregions. Each symbol corresponds to a specific subregion as defined in Figure~\ref{fig:cov_12co}. The \emph{filled points} show the smaller subregions from \citet{Ungerechts97} and \citet{Davis09}. The \emph{open points} show the larger subregions from \citet{Feddersen18}. The horizontal error bars span the cumulative lower and upper limits on $\dot P$ for the shells in each subregion (see Table 3 in \citealt{Feddersen18}).}\label{fig:slope_shell_density}
\end{figure*}

\begin{figure*}
\plotone{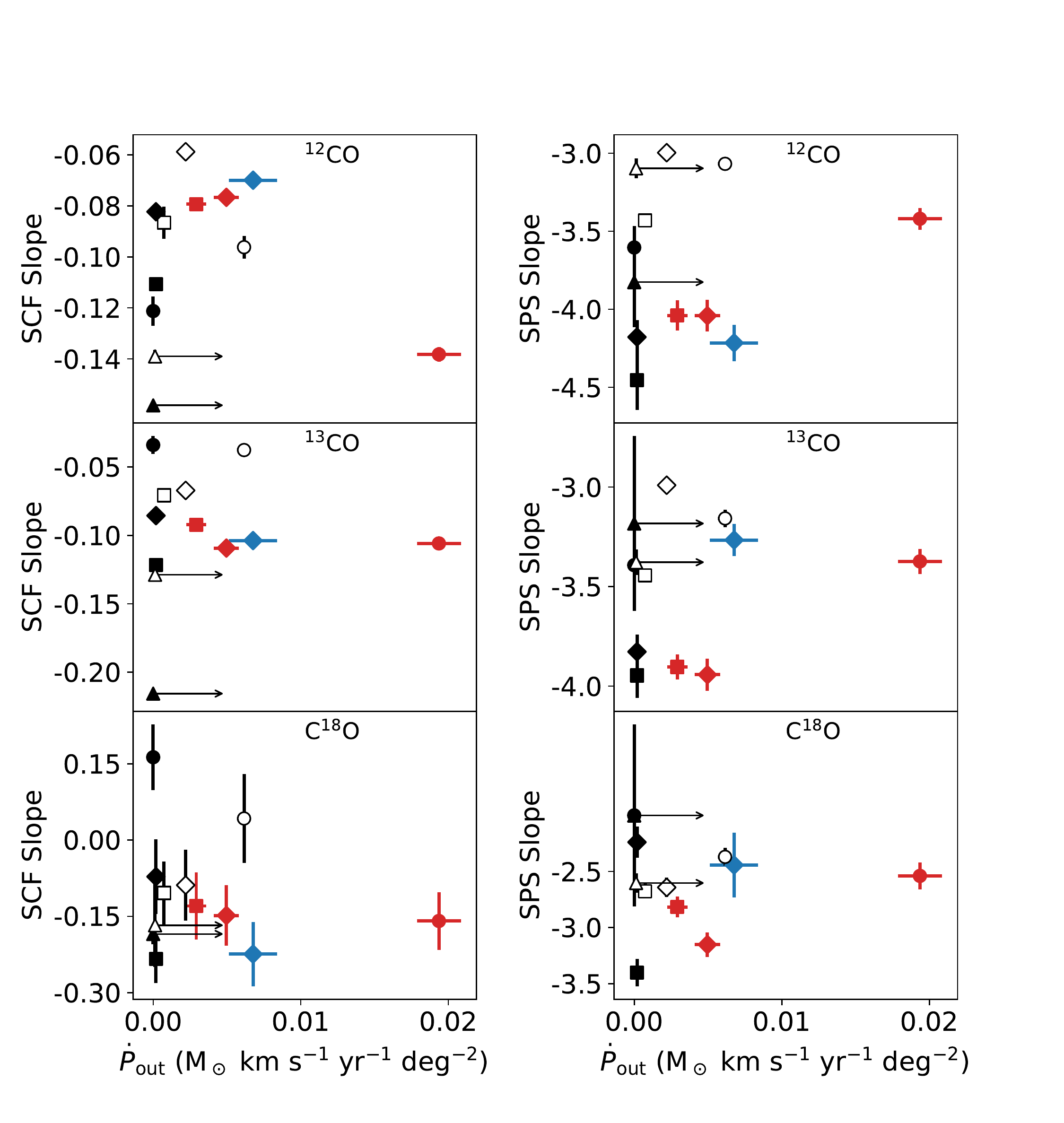}
\caption{The SCF and SPS slopes versus the momentum injection rate surface density of outflows in Orion A subregions. The arrows indicate lower limits in the regions around OMC 1, which was avoided by the outflow search of \citet{Tanabe:submitted}. All other symbols have the same meaning as in Figure~\ref{fig:slope_shell_density}. The horizontal error bars indicate the uncertainty found by adding in quadrature the individual outflow uncertainties (in Table 7 of \citealt{Tanabe:submitted}) in each subregion.}\label{fig:slope_outflow_density}
\end{figure*}

\begin{figure*}
\plotone{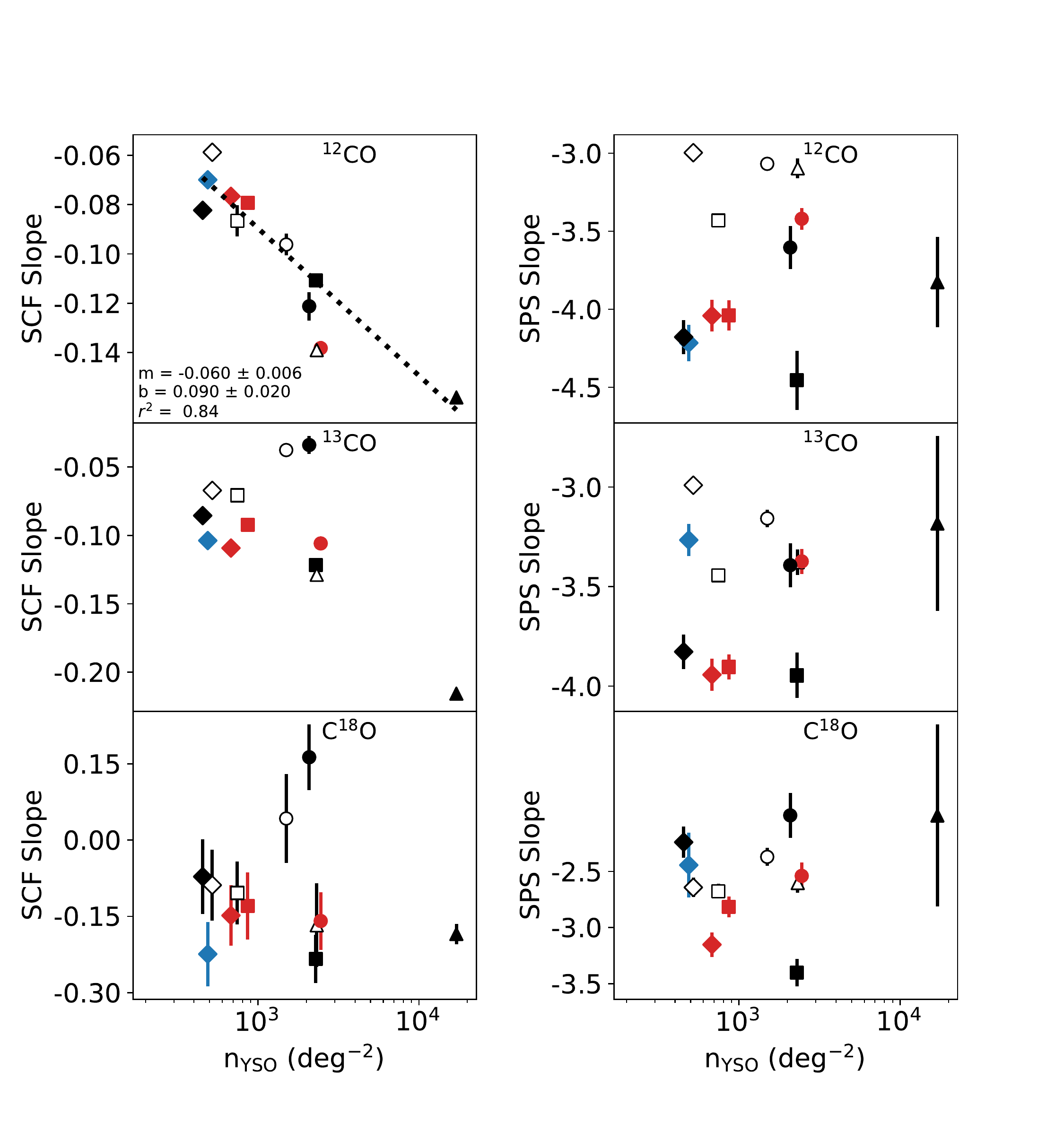}
\caption{The SCF and SPS slopes versus the surface density of YSOs in Orion A subregions. The \co[12] SCF slopes (\emph{upper left}) are fit with a linear least-squares regression (\emph{dotted line}). The slope m, intercept $b$, and correlation coefficient $r^2$ are shown. All other symbols have the same meaning as in Figure~\ref{fig:slope_shell_density}. \label{fig:slope_nstar_density}}
\end{figure*}

We quantify both SCF and SPS by the slope of their power-law fits which are shown for \co[12] in Figure~\ref{fig:scf_12co}~and~Figure~\ref{fig:sps_12co}. Figures~\ref{fig:slope_shell_density}, \ref{fig:slope_outflow_density}, and \ref{fig:slope_nstar_density} plot the SCF and SPS slopes in each subregion against the shell momentum injection rate, outflow momentum injection rate, and YSO surface densities, respectively. We look for systematic trends between our feedback impact measures and either statistic.

In our most direct comparison to \citetalias{Boyden16}, we find no correlation between the SCF or SPS slopes and shell momentum injection rate surface density in the Orion A subregions (Figure~\ref{fig:slope_shell_density}). In particular, we do not find the predicted SCF steepening with increased wind feedback. Using only those shells with the highest confidence score (i.e., a score of 5) from Table 2 in \citet{Feddersen18}, there is still no significant trend between shell momentum injection rate and SCF or SPS slope. We also find no correlation between the spectral slopes and the outflow momentum injection rate surface density (Figure~\ref{fig:slope_outflow_density}). Because both of these feedback measures are quite uncertain (see Section~\ref{sec:quantifying_feedback}), this analysis does not rule out an underlying correlation between the momentum injection rates and the spectral slopes. Using the more objective measure of YSO surface density, we do find a significant correlation with SCF slope. 

The \co[12] SCF steepens in subregions with higher YSO surface density (Figure~\ref{fig:slope_nstar_density}). While \citetalias{Boyden16} does not model YSO populations, they show that increased feedback strength steepens the \co[12] SCF in their simulations. If the true feedback impact is positively correlated with $n_{\rm{YSO}}$, then our result is consistent with the SCF prediction by \citetalias{Boyden16}. We stress that feedback is not the only possible driver of the relationship between $n_{\rm{YSO}}$ and \co[12] SCF. We tested the effects of column density on these relationships using the Herschel maps from \citet{Stutz15}. The median column density of a subregion does not affect the trends shown in Figures~\ref{fig:slope_shell_density} through \ref{fig:slope_nstar_density}. However, some other underlying variable that correlates strongly with stellar density and influences \co[12] emission (e.g., opacity or excitation temperature) may still explain the trend we see between $n_{\rm{YSO}}$ and \co[12] SCF.  

We fit the \co[12] SCF slopes with a weighted least-squares regression, shown in the upper left panel of Figure~\ref{fig:slope_nstar_density}. The best-fit line is
\begin{equation}
\begin{split}\label{eq:fit}
\alpha_{\mathrm{SCF}} = & (-0.060 \pm 0.006)~\log(n_{\mathrm{YSO}} [\mathrm{deg}^{-2}])\\ + & (0.090 \pm 0.020)
\end{split}
\end{equation}
where $\alpha_{\mathrm{SCF}}$ is the \co[12] SCF slope. This fit has a correlation coefficient of $r^2 = 0.84$. If we exclude the OMC-1 subregion, which has an order of magnitude higher $n_\mathrm{YSO}$ than any other subregion, the slope of the best-fit line steepens slightly but the correlation coefficient does not change. Future studies of the SCF in molecular clouds should test this correlation.

We do not find any correlation between SPS slope and $n_{\rm{YSO}}$, which is also consistent with \citetalias{Boyden16}. In \co[13] and C$^{18}$O, neither SCF nor SPS appear to be correlated with $n_{\rm{YSO}}$. It is unclear why the \co[12] SCF slope is correlated with $n_{\rm{YSO}}$ while the other CO lines are not. 

When measuring the SCF slopes, we fit the SCF up to about 0.12 pc scales. We chose this fitting range to be most consistent with \citetalias{Boyden16}. The SCF steepens toward larger scales. However, fitting the SCF at larger scales does not change the relative trends between SCF slope and feedback impact reported in Figures~\ref{fig:slope_shell_density} through~\ref{fig:slope_nstar_density}. In particular, the anticorrelation between \co[12] SCF slope and $n_{\rm{YSO}}$ remains. We show the SCF at larger scales in Appendix~\ref{sec:appendix_scf_largescale}. 

Figure~\ref{fig:feedback_three} shows the three feedback measures plotted against each other. There is no significant correlation between any of the three feedback measures among the Orion A subregions. This lack of correlation is unsurprising. As discussed in Section~\ref{sec:quantifying_feedback}, shells, outflows, and YSOs each trace different populations of forming stars. For this reason, along with the significant uncertainties in the feedback measures also described in Section~\ref{sec:quantifying_feedback}, it is premature to conclude that the feedback mechanisms are not spatially correlated.

\begin{figure*}
\plotone{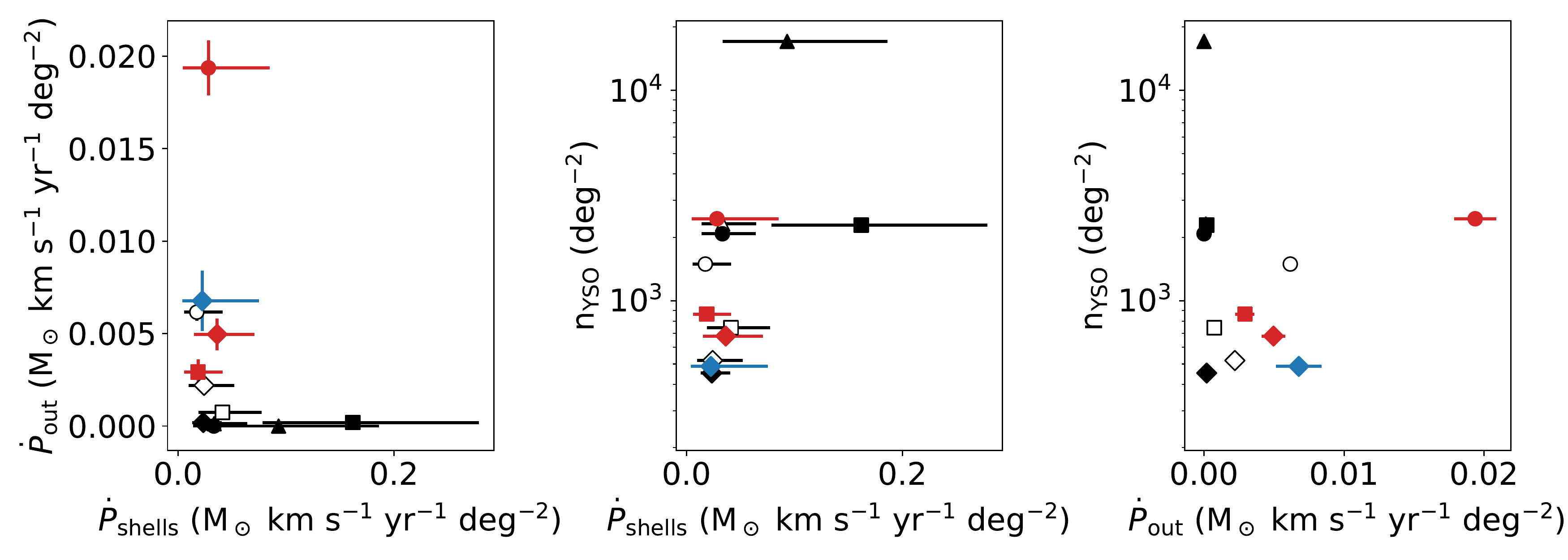}
\caption{The three feedback measures plotted against one another. The symbols have the same meaning as in Figure~\ref{fig:slope_shell_density}. There is no significant correlation between any of the measures.  \label{fig:feedback_three}}
\end{figure*}

\subsection{Line Wing Power Spectrum}\label{sec:discussion_linewing}
The power spectra in Figure~\ref{fig:sps_12co} are calculated using the \co[12] integrated intensity maps. Because outflows are often most prominent in the line wings (channels blueward and redward of the main cloud velocity range), their influence may be largest in power spectra restricted to these velocity ranges. 

\citet{Swift08} studied the line wing power spectrum of the low-mass star-forming cloud L1551. They found a peak at an angular scale of 1\arcmin~(0.05 pc) in the power spectrum of \co[13] integrated over the line wings. They attributed this feature to a preferential scale at which protostellar outflows deposit energy into the cloud.

\citet{Padoan09} showed that the integrated intensity power spectrum in the NGC 1333 molecular cloud is nearly a perfect power law with no features or flattening despite the many outflows present \citep{Plunkett13}. \citet{Padoan09} interpret this to mean that all turbulent driving takes place at large scales, presumably by external driving forces, and that outflows do not drive enough momentum into the cloud to affect the turbulent cascade. However, \citet{Padoan09} focused on the power spectrum of integrated intensity maps, while \citet{Swift08} specifically examined the power spectra in the line wings.

\begin{figure*}
\plotone{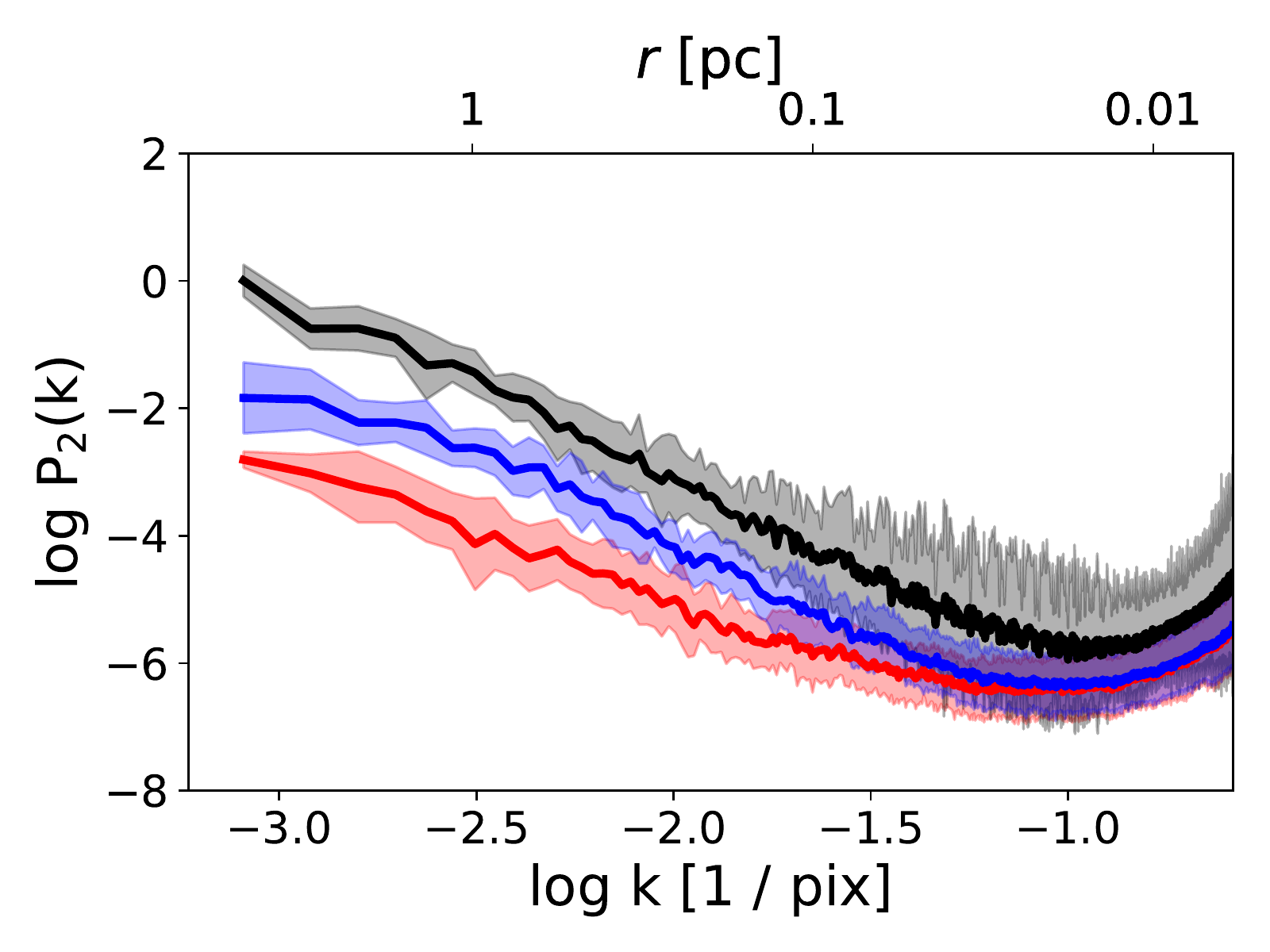}
\caption{\co[13] line wing power spectra in L1641N subregion. The \emph{blue line} shows the blue line wing power spectrum of emission integrated from 4.3 to 5.6 \kms. The \emph{red line} shows the red line wing power spectrum of emission integrated from 10.8 to 12.1 \kms. The \emph{black line} shows the power spectrum at main cloud velocities 6.6 to 7.8 \kms. The spectra are divided by the first value of the \emph{black} power spectrum for comparison purposes. Each power spectrum is beam corrected and tapered with a Tukey window. The upturn at log k = -1 pix$^{-1}$ is a numerical effect caused by the beam correction.\label{fig:sps_linewing_13co}}
\end{figure*}

To investigate the line wing power spectra in Orion A, we visually define the velocity ranges of the line wings where the main cloud emission disappears. Then, we integrate the emission in these velocity ranges and compare their power spectra. Figure~\ref{fig:sps_linewing_13co} shows the \co[13] line wing power spectrum toward the L1641N subregion defined in \citet{Feddersen18}, which contains several outflows. In L1641N, we define the blueshifted line wing as 4.3 to 5.6 \kms, the central cloud velocity range as 6.6 to 7.8 \kms, and the redshifted line wing as 10.8 to 12.1 \kms. We find no sign of peaks in the \co[12] or \co[13] line wing power spectra of any subregion in Orion A.

While the \co[12] and \co[13] power spectra are featureless power-laws, we do find peaks in the \co[18] power spectrum at 30-60\arcsec. In Appendix~\ref{sec:pspec_noise}, we show that these peaks are present in the power spectrum of emission-free channels and are therefore artifacts of the data combination process.

We suggest that the peak in the \co[13] line wing power spectra found by \citet{Swift08} may be caused by a similar numerical artifact in their data. This peak occurs over angular scales of approximately 40 to 100\arcsec. \citet{Swift08} combined an interferometric mosaic from the Berkeley-Illinois-Maryland Association (BIMA) array with a map from the Arizona Radio Observatory (ARO) 12m telescope. The spacing between BIMA pointings is 45\arcsec~and the FWHM of the ARO 12m beam is 55\arcsec. These angular scales are similar to scale of the peak in the power spectrum, suggesting that an artifact introduced in the \co[13] data reduction or combination may be responsible for this feature. To test if the peak is real or a data artifact, the power spectrum of emission-free channels should be compared. If the peak appears in this noise power spectrum, it is not intrinsic to the emission and does not indicate a characteristic scale imposed by outflows. We suggest that future studies of molecular cloud power spectra first consider the noise power spectrum before interpreting any deviations from a smooth power-law.

\section{Conclusions}\label{sec:conclusions}

This study is one of the first attempts to quantify the connection between feedback and the statistics of turbulent motion in a molecular cloud. Previous studies of feedback in molecular clouds have focused on cataloging and measuring individual features like outflows and shells. 

We find spatially correlated emission at relative velocities of 1-3~\kms~in the covariance matrix toward the NGC 1977, OMC4, L1641N, and V380 regions. These features resemble those found by \citetalias{Boyden16} in their simulations of stellar winds. It is unclear whether winds are responsible for these features in Orion, or whether they arise from some other mechanism, such as the cloud-cloud collision proposed for L1641-N by \citet{Nakamura12}. We suggest a detailed comparison of the covariance matrix in simulations that incorporate both feedback and a colliding-cloud model to clarify what mechanism produces these features in Orion A. 

Contrary to the predictions of the \citetalias{Boyden16} simulations, we do not find a relationship between the slope of the spectral correlation function in Orion A and the momentum injection rate of shells or outflows. However, the uncertainties inherent in both shell and, to a lesser extent, outflow momenta mean we cannot rule out an underlying relationship. A better accounting of the impact of these feedback processes is necessary to fully understand their relationship to the statistics of turbulence. 

We find, for the first time, a significant trend between the spectral correlation function and the surface density of young stars. Regions with higher YSO surface density have steeper spectral correlation functions in \co[12]. If higher YSO surface density correlates with greater feedback impact, the \co[12] SCF slope may be a useful indicator of the importance of feedback in molecular clouds. Feedback is not the only possible underlying variable in this relationship. Any parameter that correlates strongly with stellar density (such as optical depth or excitation temperature) may contribute to the trend we see with the \co[12] SCF.

We find no significant trend between the power spectrum in Orion A and the momentum injection rate of shells or outflows, in agreement with \citetalias{Boyden16}. We find no evidence for features or breaks in the integrated intensity power spectra or  line wing power spectra. Thus, we find no evidence in Orion A for a preferential scale imposed by feedback. Future studies of molecular cloud power spectra should first examine the power spectrum of noise before interpreting any deviations from a smooth power-law.

The statistical study of feedback in molecular clouds is still in its infancy. Future simulations of molecular clouds should compare statistical measures of gas structure and kinematics with the \textit{strength} of feedback, beyond merely its presence or absence. To aid this, simulations should explore parameter space more fully \citep[see e.g.][]{Yeremi14} to disentangle the impacts of feedback from other factors like chemistry and magnetic fields. Future studies should also differentiate between the statistical effects of protostellar outflows, spherical stellar winds, and other types of feedback.

\acknowledgments 
We thank Yoshihiro Tanabe for providing the outflow catalogue. We thank Jens Kauffmann, Jaime Pineda, and the entire CARMA-NRO Orion collaboration for helpful discussion. We thank the referee for thoughtful suggestions that greatly improved the paper. CARMA operations were supported by the California Institute of Technology, the University of California-Berkeley, the University of Illinois at Urbana-Champaign, the University of Maryland College Park, and the University of Chicago. The Nobeyama 45 m telescope is operated by the Nobeyama Radio Observatory, a branch of the National Astronomical Observatory of Japan. 
This project was supported by the National Science Foundation, award AST-1140063.

\facilities{CARMA, No:45m}
\software{TurbuStat \citep{Koch17}}

\appendix
\section{Statistics of \co[13] and C$^{18}$O}\label{sec:appendix_13co_c18o}

\subsection{Covariance Matrix}
\begin{figure}
 \gridline{\fig{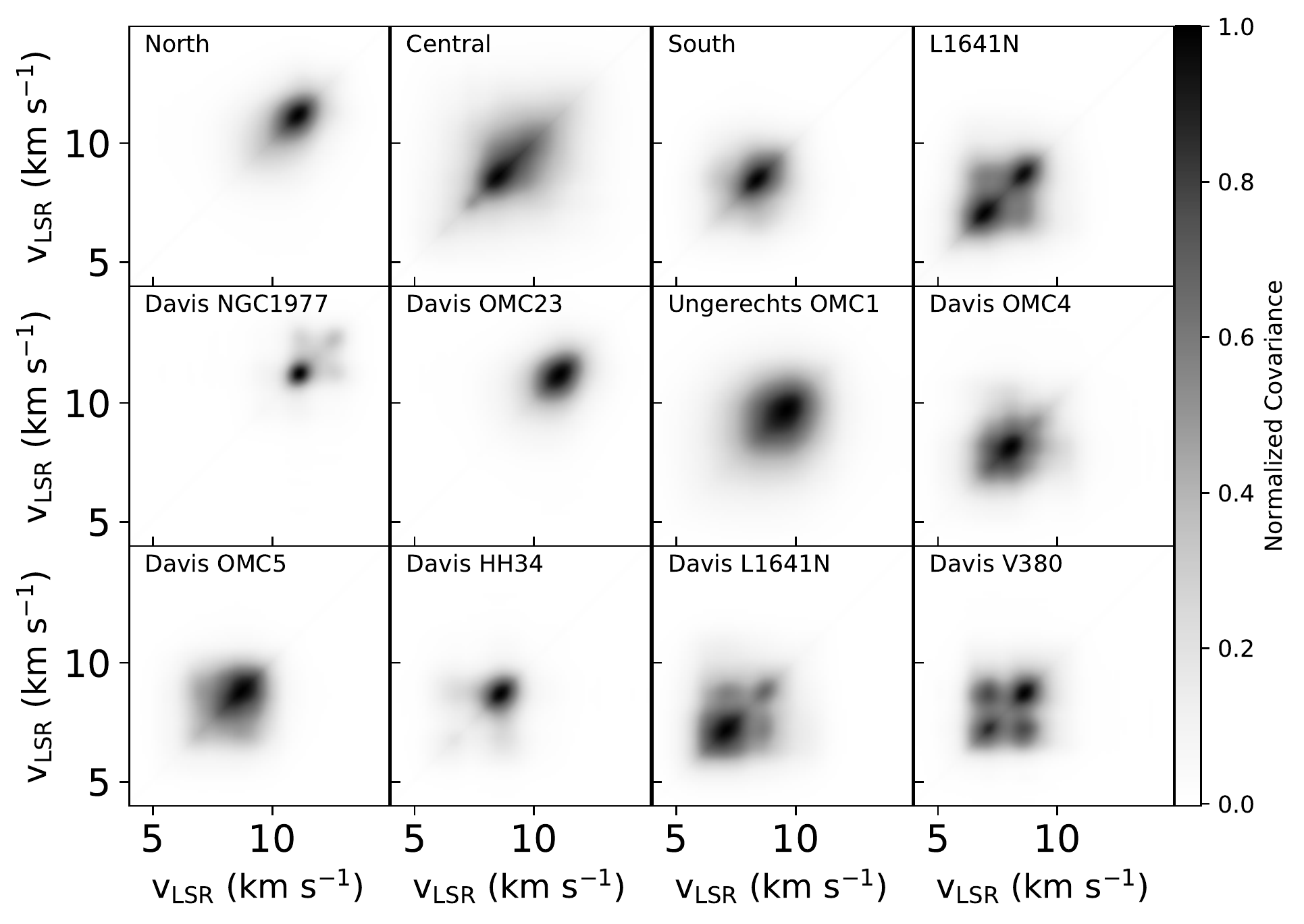}{0.8\textwidth}{(a)}}
\gridline{\fig{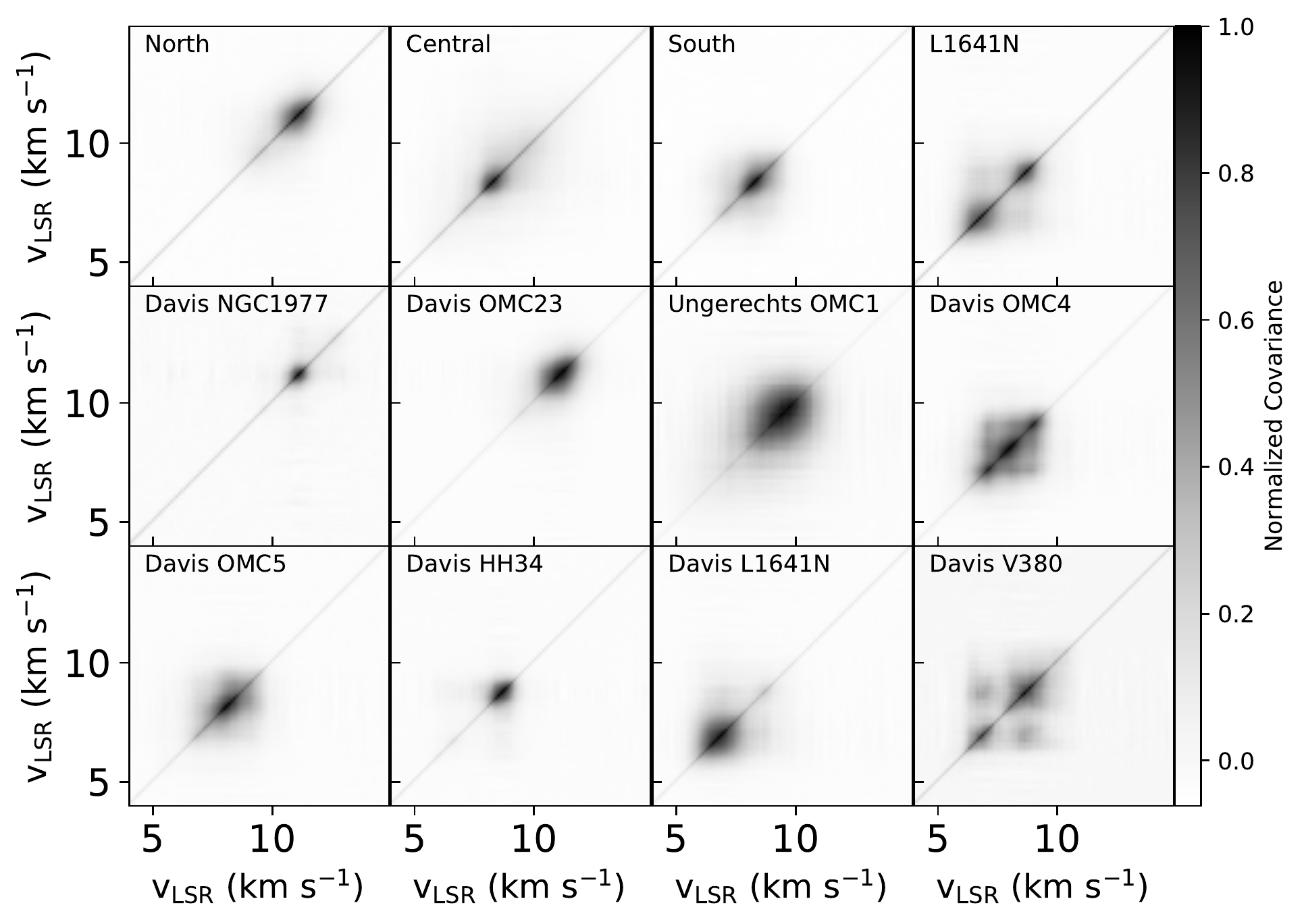}{0.8\textwidth}{(b)}}
\caption{(\emph{a}) \co[13] and (\emph{b}) C$^{18}$O covariance matrices in Orion A subregions.\label{fig:appendix_cov_13co_c18o}}
\end{figure}

We show the \co[13] and C$^{18}$O covariance matrices in Figure~\ref{fig:appendix_cov_13co_c18o}. The off-axis peaks seen in the \co[12] covariance matrices are also present in \co[13] and C$^{18}$O. Because these features are prominent in both optically thick and thin tracers, they likely result from real velocity structure in the cloud rather than simply from optical depth effects. \citetalias{Boyden16} computed the covariance matrix of their simulated cube before modeling radiative transfer (Figure 21 in \citetalias{Boyden16}). This pre-processed cube does not show covariance peaks, implying that excitation and optical depth effects enhance this feature of their winds. The mapping between these pre-processed cubes and observed optically thin CO lines is unclear. We suggest future models incorporate multiple observable transitions to better compare modeled and observed covariance.

\subsection{SCF}
\begin{figure}
 \gridline{\fig{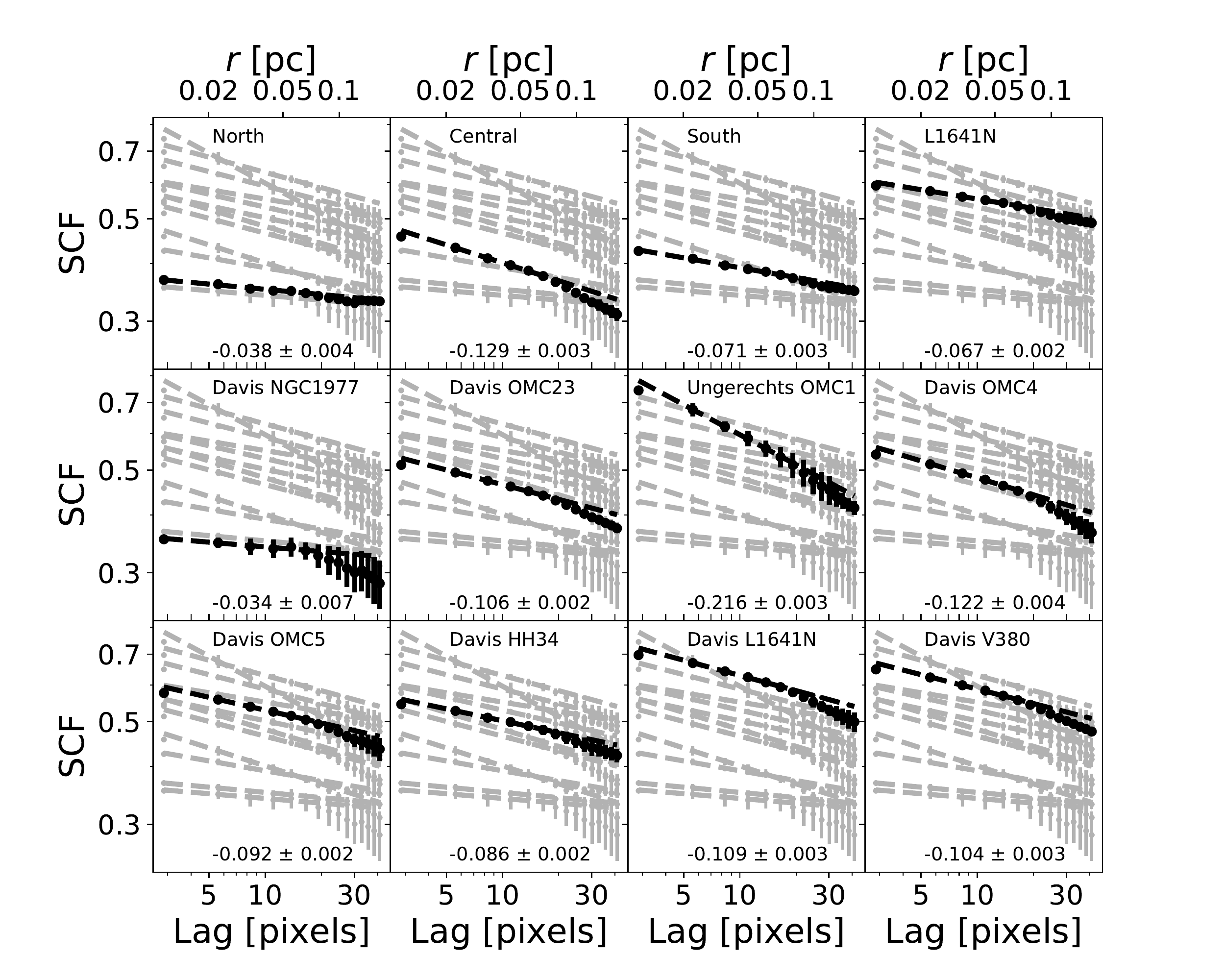}{0.7\textwidth}{(a)}}
\gridline{\fig{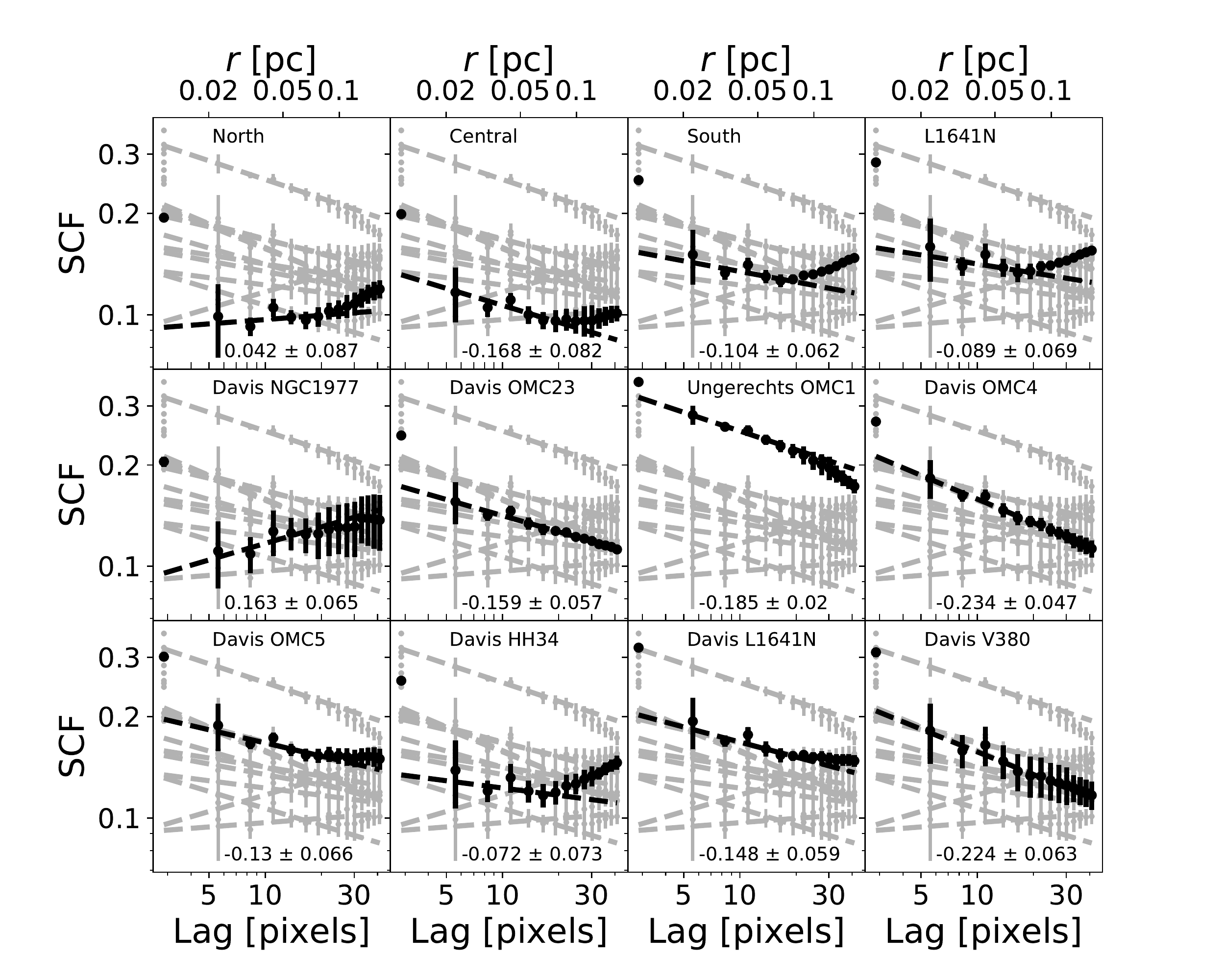}{0.7\textwidth}{(b)}}
\caption{(\emph{a}) \co[13] and (\emph{b}) C$^{18}$O spectral correlation function and power-law fits in Orion A subregions. The symbols and lines have the same meaning as in Figure~\ref{fig:scf_12co}.\label{fig:appendix_scf}}
\end{figure}

We show the \co[13] and C$^{18}$O SCF in Figure~\ref{fig:appendix_scf} and the \co[13] and C$^{18}$O SPS in Figure~\ref{fig:appendix_sps}. The \co[13] SCF spectra have shapes similar to the \co[12] spectra. They follow power-laws up to lags of about 20 pixels ($40\arcsec$ or $0.08$~pc), where they steepen toward larger scales. 

The C$^{18}$O SCF spectra look very different. They show a sharp decrease between the shortest lags of 3 and 5 pixels (6-10$\arcsec$ or 0.01-0.02 pc). Many of the C$^{18}$O SCF spectra also turn upward at larger scales. An upward sloping SCF means that distantly separated spectra are more similar than close pairs of spectra, an unphysical result. Because the C$^{18}$O has low signal-to-noise, the upward slopes are more likely to be an artifact of the data reduction process rather than a real signature of the cloud spectra.

\subsection{SPS}
\begin{figure}
\gridline{\fig{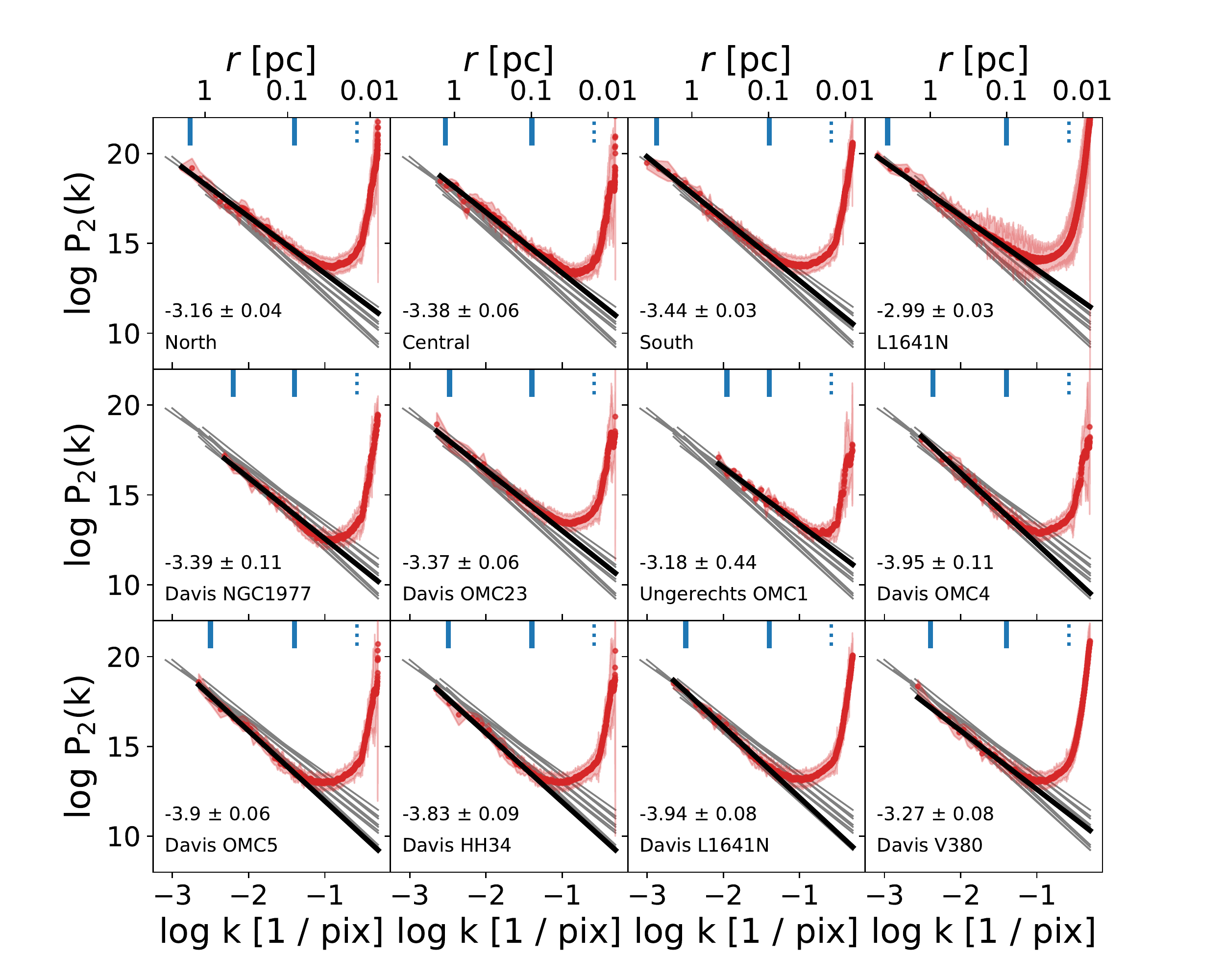}{0.7\textwidth}{(a)}}
\gridline{\fig{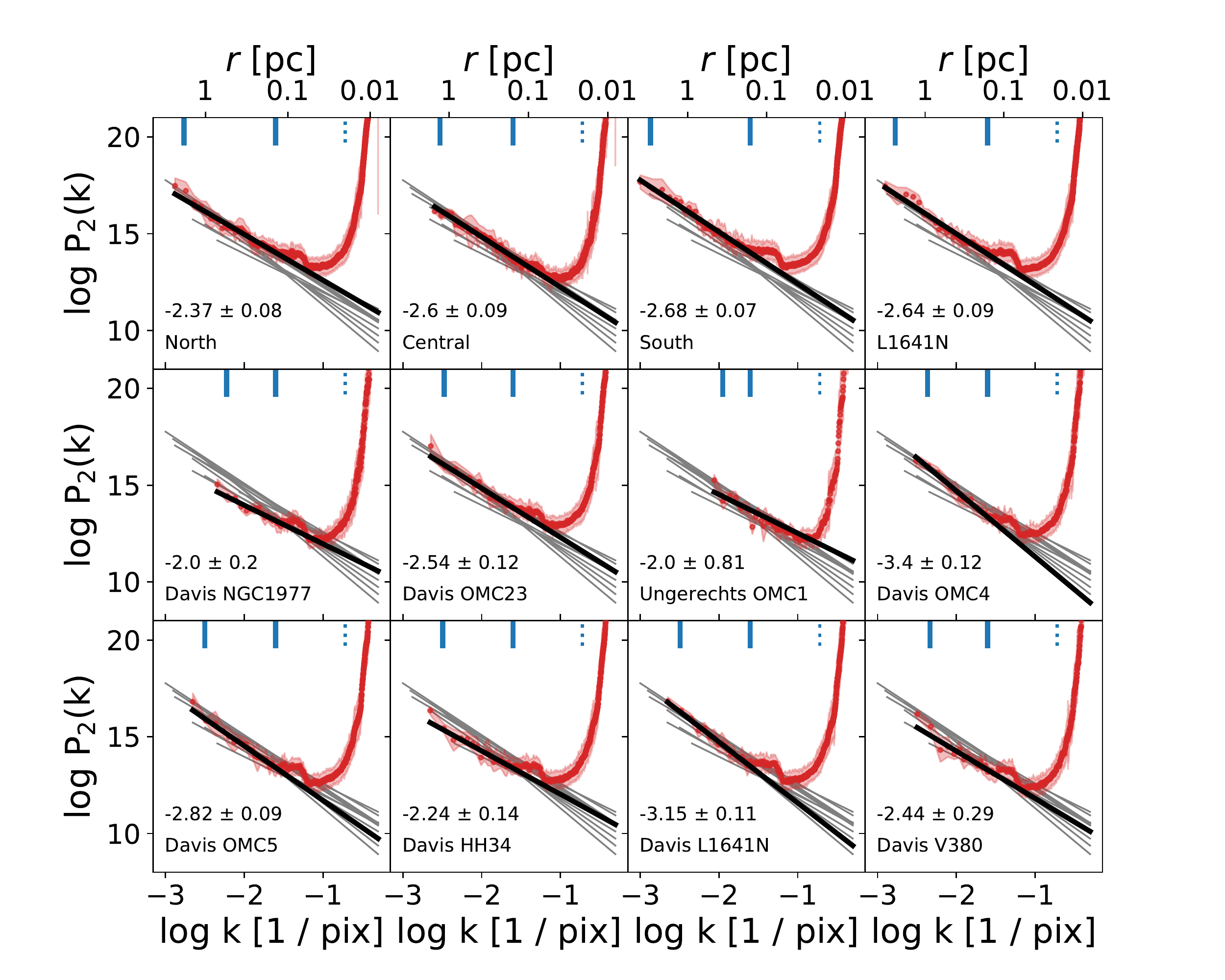}{0.7\textwidth}{(b)}}
\caption{(\emph{a}) \co[13] and (\emph{b}) C$^{18}$O spatial power spectrum and power-law fits in Orion A subregions. The symbols and lines have the same meaning as in Figure~\ref{fig:sps_12co}.
The bump at 0.1 pc in the C$^{18}$O power spectrum,  an artifact of the data, is evident in most panels (see 
\S~\ref{sec:pspec_noise})
\label{fig:appendix_sps}}
\end{figure}

As in \co[12], the \co[13] SPS (Figure~\ref{fig:appendix_sps}) closely follows a power-law down to scales of a few beamwidths. The slopes of the \co[13] SPS fits range from -3 to -4. Overall, the \co[13] slopes are slightly shallower than \co[12], closer to the predicted value of -3 for an optically thick medium and the observational results compiled by \citet{Burkhart13}, although Section 5.2 in \citet{Kong18} indicates that \co[13] is not very optically thick in Orion A, with only 0.6\% of pixels having $\tau_{^{13}\mathrm{CO}} > 1$.

The C$^{18}$O SPS have significantly shallower slopes than \co[12] or \co[13]. However, they show a clear peak near 0.1 pc. This peak remains in the power spectrum of emission-free channels (see Appendix~\ref{sec:pspec_noise}). Because of the low signal-to-noise of C$^{18}$O, we do not interpret these power spectra. 

\section{Noise Peak in C$^{18}$O SPS}\label{sec:pspec_noise}
The C$^{18}$O SPS has a peak near 0.1~pc which is not present in either \co[12] or \co[13]. This feature can also be seen in the integrated C$^{18}$O delta-variance spectra in Figure 3 of \citet{Kong18}, where they speculate it arises from the low signal-to-noise of C$^{18}$O coupled with problems in the map cleaning process. \citet{Swift08} found a similar feature in the \co[13] line-wing power spectrum of the L1551 cloud and attributed this SPS feature to feedback (see Section~\ref{sec:discussion_linewing}). To rule out the influence of feedback in our C$^{18}$O data, we examine the noise power spectrum. If the feature at 0.1~pc is present in the noise, it is not related to feedback. 

\begin{figure*}
\plotone{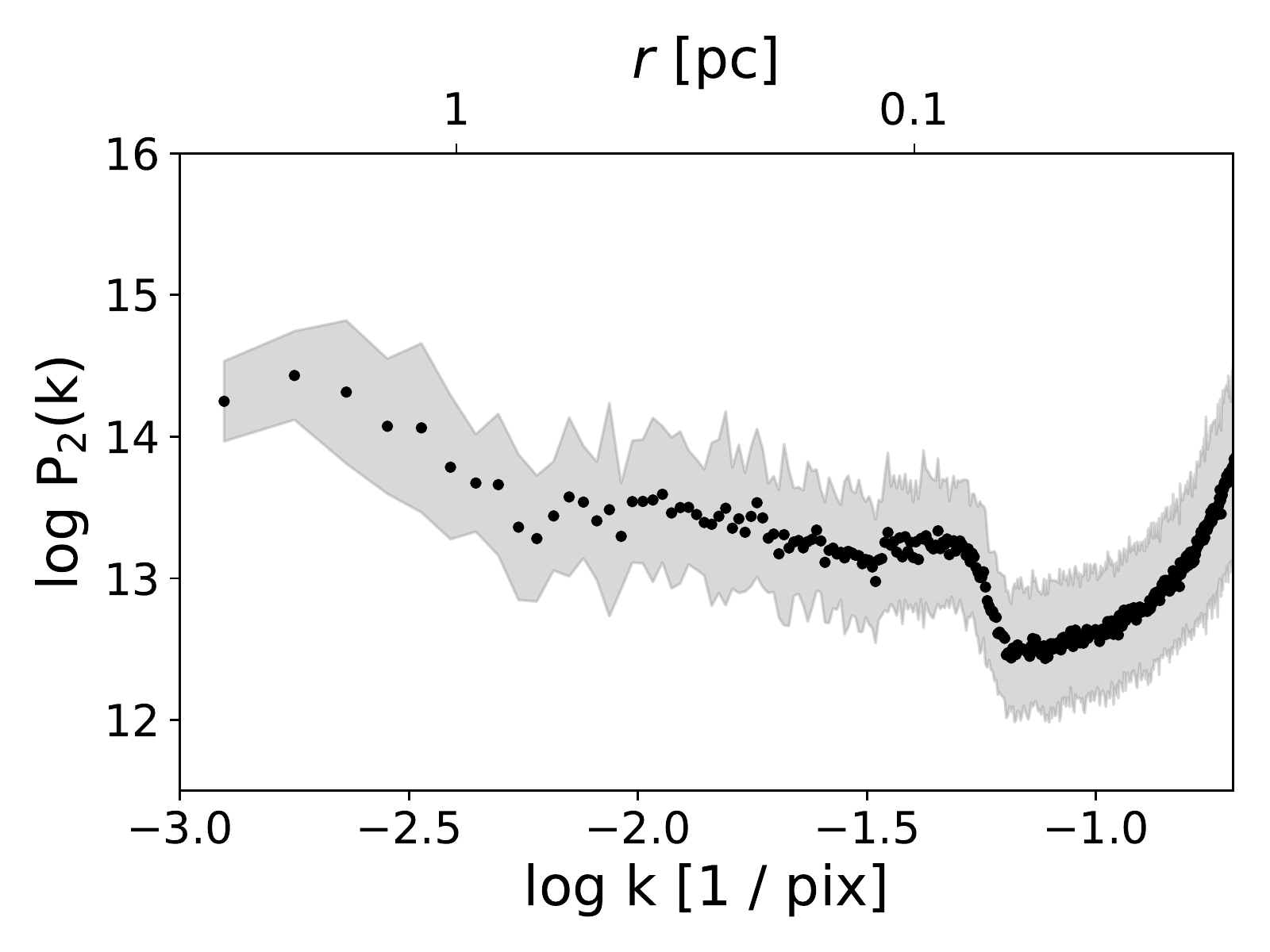}
\caption{The C$^{18}$O noise power spectrum toward L1641N. The power spectrum is calculated in the emission-free channels between 12 and 16~\kms. The bump at 0.1~pc is an artifact of the data combination process and also appears in the C$^{18}$O delta-variance spectrum shown by \citet{Kong18}.\label{fig:appendix_sps_noise_c18o}}
\end{figure*}

To compute the noise power spectrum of C$^{18}$O, we first sum the emission-free channels between 12 and 16~\kms. Then we compute the spatial power spectrum as described in Section~\ref{sec:methods_sps}, including the beam correction and tapering. We show the power spectrum of noise in the L1641N region C$^{18}$O map in Figure~\ref{fig:appendix_sps_noise_c18o}. The noise shows the 0.1~pc peak. Because the C$^{18}$O map has much lower signal-to-noise than \co[12] and \co[13], the shape of its noise power spectrum dominates the power spectrum of integrated emission. While the \co[12] and \co[13] noise power spectra also deviate from power-law, their higher signal-to-noise renders the noise power spectrum insignificant. Before interpreting features in the SPS of any new dataset, we suggest calculating the noise power spectrum. 

\section{SCF at Large Scales}\label{sec:appendix_scf_largescale}
\begin{figure*}
\plotone{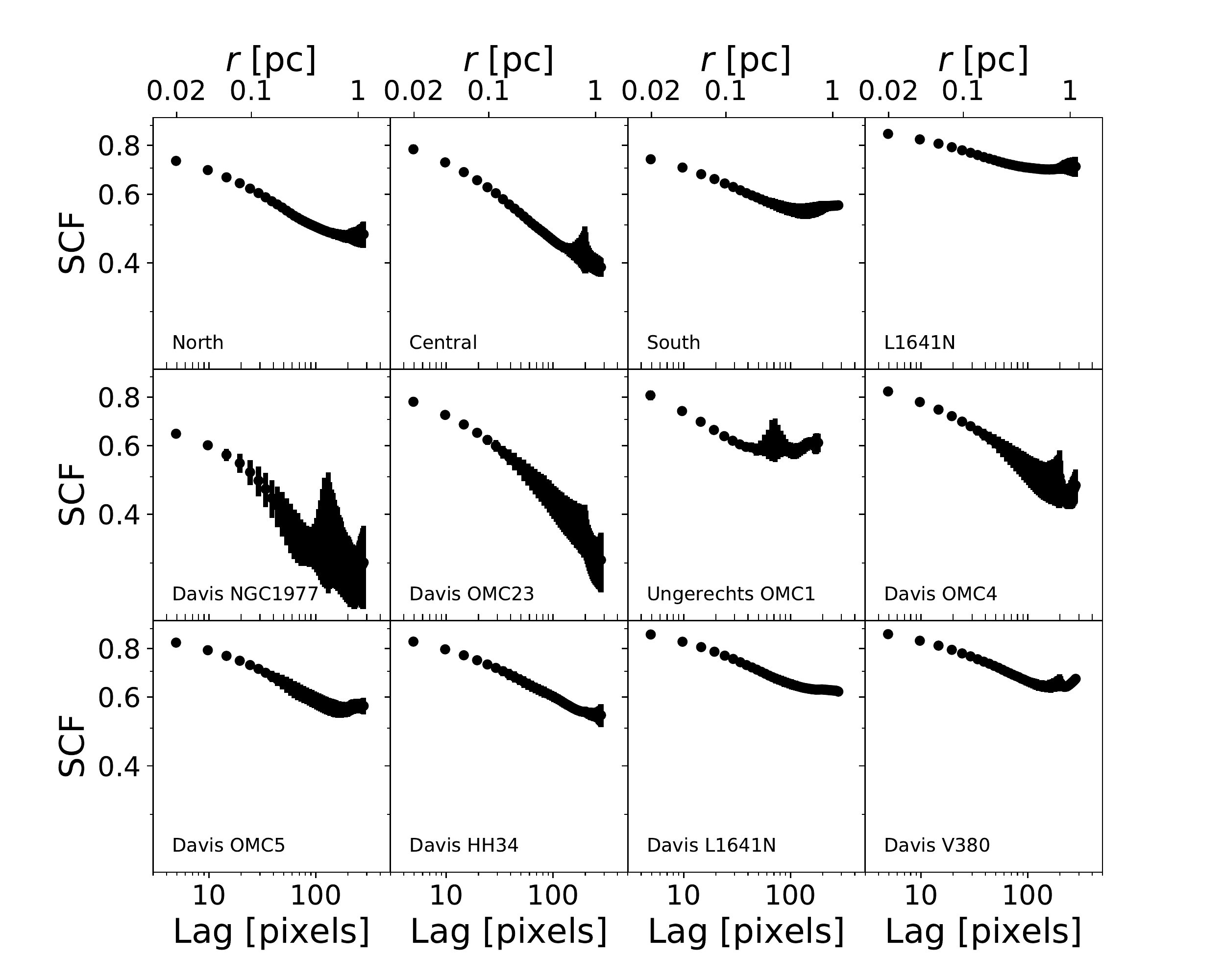}
\caption{The \co[12] large-scale spectral correlation function in Orion A subregions.\label{fig:appendix_scf_largescale}}
\end{figure*}

The SCF shown in Figure~\ref{fig:scf_12co} only extends to lags of about 45 pixels, or 0.17~pc, to match the range in physical scales covered by the SCF in \citetalias{Boyden16}. However, other observational studies fit the SCF spectra at larger scales \citep[e.g.][]{Padoan03,Gaches15}. In Figure~\ref{fig:appendix_scf_largescale}, we show the \co[12] SCF spectra of the Orion A subregions at scales up to 300 pixels (1.2 pc). The SCF steepens at larger scales, in closer agreement with the SCF measured by \citet{Padoan03} and \citet{Gaches15}. Each subregion SCF flattens and turns upward at different lags. This shape at large scales is a numerical effect caused by the finite size of the subregions and makes it difficult to compare the SCF at large scales between maps of different sizes. If we fit each subregion SCF with a power-law between 30 and 50 pixels (0.12 to 0.2 pc), which avoids the SCF upturn for all subregions, the relationship between SCF slope and feedback impact remains qualitatively the same as shown in Figures~\ref{fig:slope_shell_density}~through~\ref{fig:slope_nstar_density}.

\section{Full-Map Statistics}\label{sec:fullmap}
\begin{figure*}[h!]
\plotone{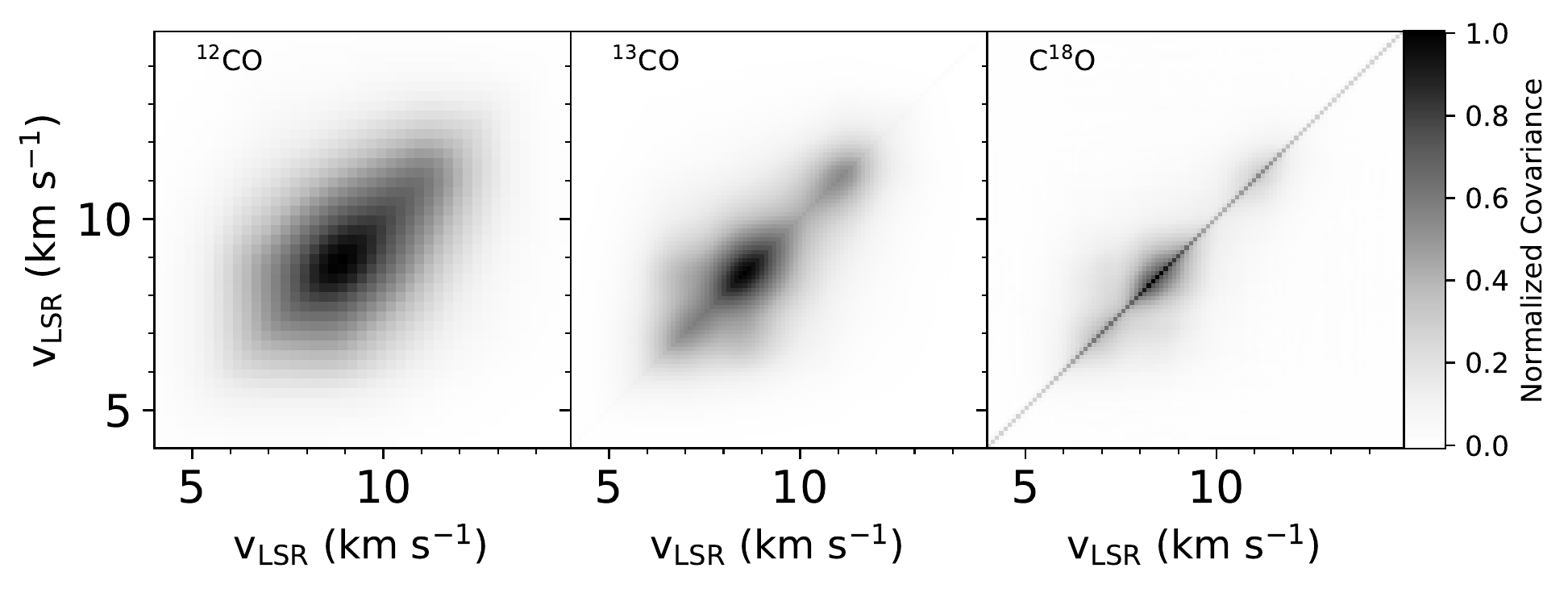}
\caption{The covariance matrices of the full Orion A map. \label{fig:appendix_cov_fullmap}}
\end{figure*}

\begin{figure*}[h!]
\epsscale{0.9}
\plotone{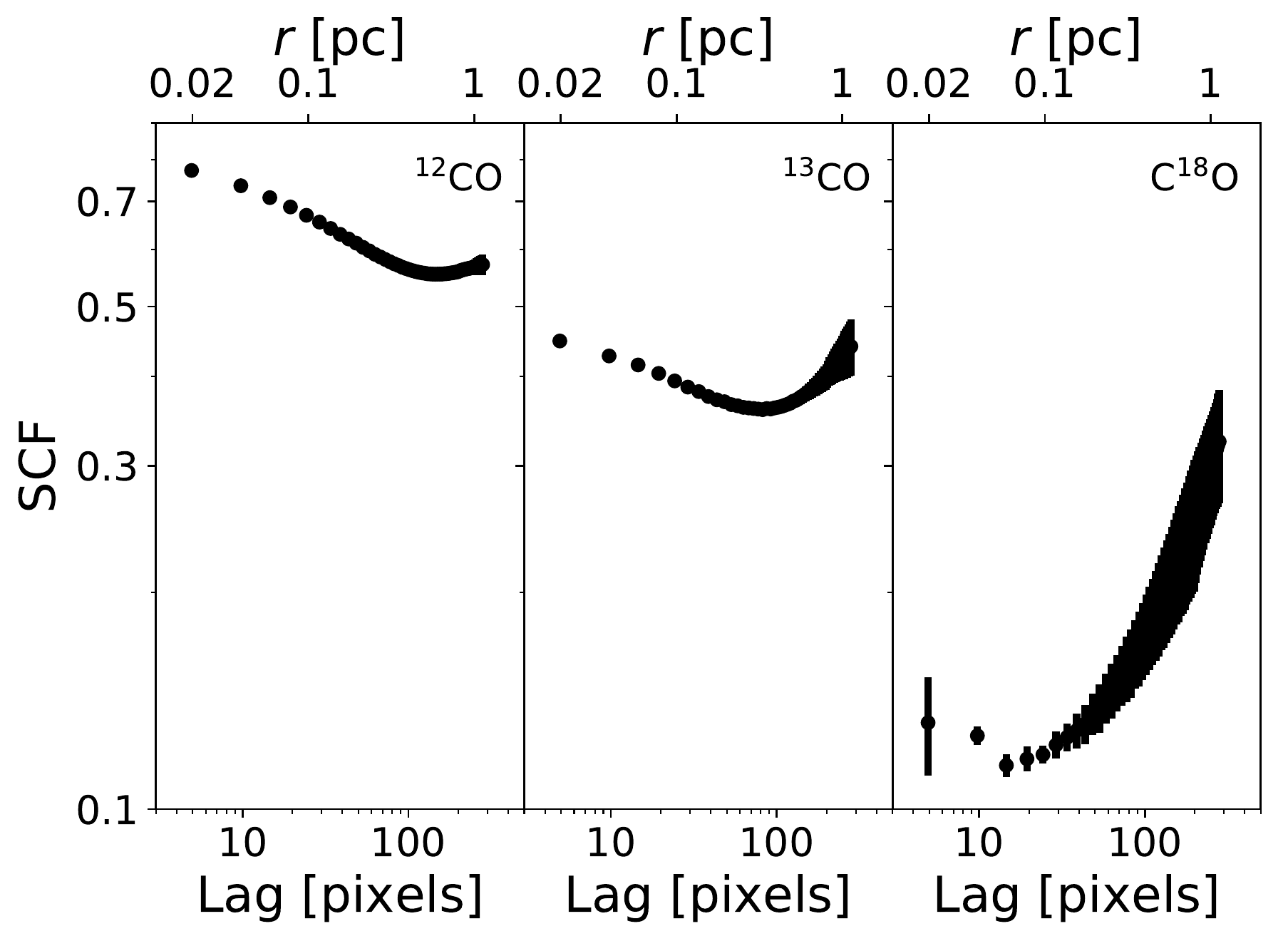}
\caption{The spectral correlation function of the full Orion A map at large scales as in Figure~\ref{fig:appendix_scf_largescale}.\label{fig:appendix_scf_fullmap}}
\end{figure*}

\begin{figure*}[hb!]
\epsscale{0.9}
\plotone{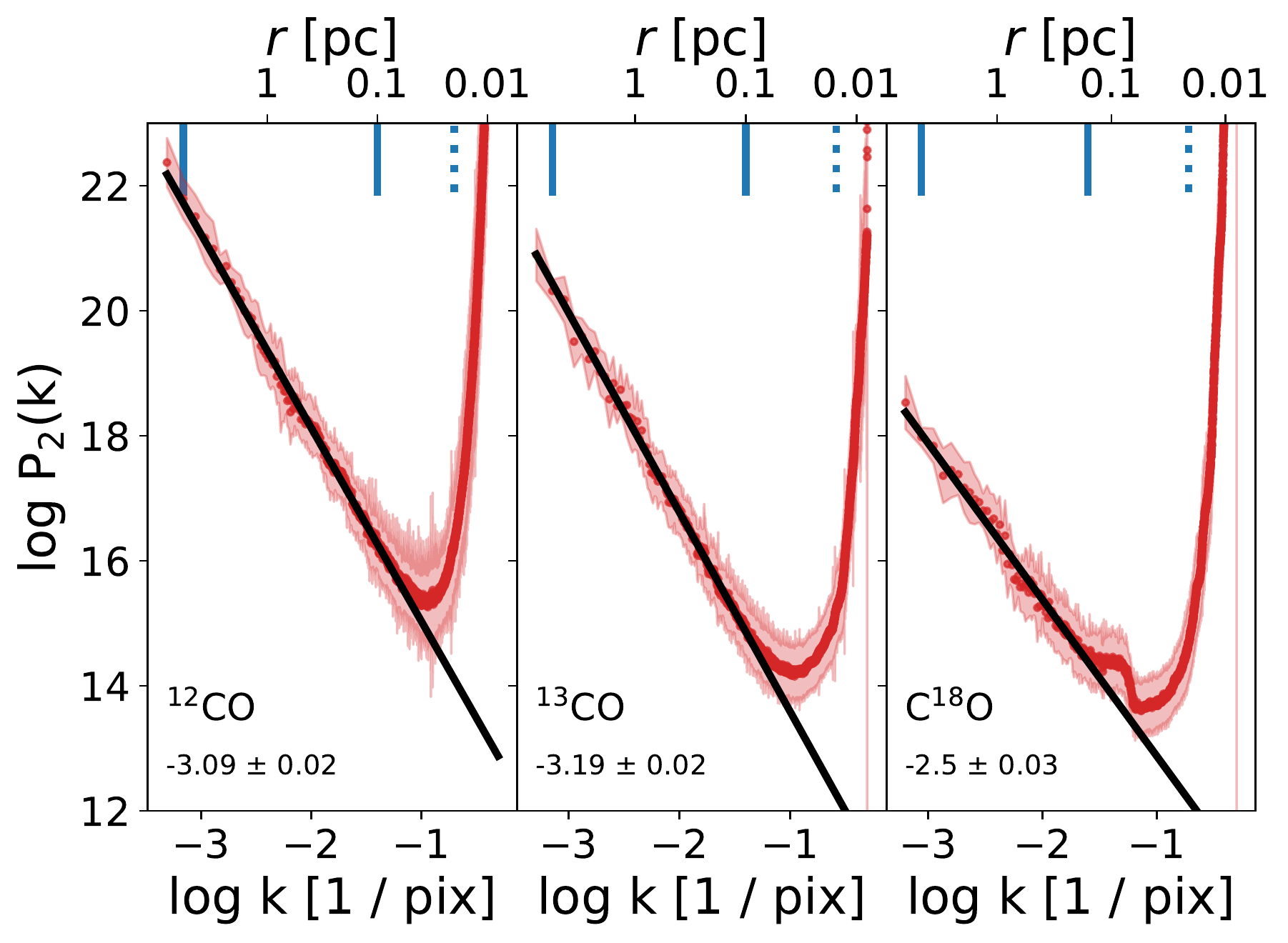}
\caption{The spatial power spectrum and power-law fits of the full Orion A map. All symbols and lines have the same meaning as in Figure~\ref{fig:sps_12co}.
The bump at 0.1 pc in the C$^{18}$O power spectrum, is an artifact of the data combination process, also seen in Figure~\ref{fig:appendix_sps_noise_c18o}. \label{fig:appendix_sps_fullmap}}
\end{figure*}

We show the covariance matrices, SCF, and SPS computed over the entire area covered by CARMA-NRO Orion maps in Figures~\ref{fig:appendix_cov_fullmap}, \ref{fig:appendix_scf_fullmap}, and \ref{fig:appendix_sps_fullmap}, respectively.

\clearpage
\bibliographystyle{aasjournal}
\bibliography{all.bib}

\end{document}